\documentclass[useAMS,usenatbib]{mn2e}
\usepackage{graphicx,subfigure}
\usepackage{enumerate}
\usepackage{graphicx,subfigure,amsmath,wasysym, amssymb}
\title[Polar Ring Galaxies in the Galaxy Zoo]{Polar Ring Galaxies in the Galaxy Zoo}

\author[Ido Finkelman et al.]{Ido Finkelman$^{1}$\thanks{E-mail: ido@wise.tau.ac.il (IF)}, Jos\'{e} G. Funes S.J.$^{2}$, Noah Brosch$^{1}$\\
$^{1}$The Wise Observatory and the Raymond and  Beverly Sackler School of Physics and
Astronomy, the Faculty of Exact
Sciences, \\ Tel Aviv University, Tel Aviv 69978, Israel\\
$^{2}$Vatican Observatory, V-00120 Vatican City State, Italy
}

\begin{document}

\date{Accepted 2012 February 21.  Received 2012 February 21; in original form 2011 June 13}

\pagerange{\pageref{firstpage}--\pageref{lastpage}} \pubyear{2002}

\maketitle

\label{firstpage}

\begin{abstract}
We report observations of 16 candidate polar ring galaxies (PRGs) identified by the Galaxy Zoo project in the Sloan Digital Sky Survey (SDSS) database. Deep images of five galaxies are available in the SDSS Stripe82 database, while to reach similar depth we observed the remaining galaxies with the 1.8-m Vatican Advanced Technology Telescope.
We derive integrated magnitudes and $u-r$ colours for the host and ring components and show continuum-subtracted H$\alpha$+[NII] images for seven objects. 
We present a basic morphological and environmental analysis of the galaxies and discuss their properties in comparison with other types of early-type galaxies.
Follow-up photometric and spectroscopic observations will allow a kinematic confirmation of the nature of these systems and a more detailed analysis of their stellar populations.

\end{abstract}


\section{Introduction}
Polar-ring galaxies (PRGs) are one of the more remarkable objects of the large family of non-barred ring
galaxies. They are generally composed of an S0-like galaxy and a polar ring, which remain separate for
billions of years (Whitmore et al.\ 1990). 
PRGs are ideal laboratories to gain insight into galaxy formation processes because of the high inclination of their outer structures and the evidence of very recent gas infall. The mechanism by which these quasi-stable systems have formed is generally attributed
to galaxy-galaxy interactions but most of the details have not been finalized (Bekki 1998; Bournaud \& Combes 2003).

A straightforward observational approach to detect PRGs is to search for distinctive features.
The central early-type galaxy is mostly devoid of gas and is surrounded by a highly inclined ring of gas, dust and stars.
Moreover, the system needs to be in quasi-equilibrium, with the centres of the two components closely aligned. 
The first large survey aimed to identify PRGs was performed by Whitmore et al.\ (1990). Their photographic atlas of PRGs and related objects lists 157 objects, mainly selected based on their peculiar appearance.
Unfortunately, only a few kinematic follow-up studies of these objects were performed, confirming the nature of about a couple of dozens of these systems.    

Polar rings span a wide range of size and luminosity, not necessarily comparable to that of their central hosts (Reshetnikov \& Sotnikova 1997). Extended, luminous polar structures are relatively simple to detect, while faint rings are more difficult to recognize by visual inspection of traditional images of nearby galaxies, or during a serendipitous search in large and shallow surveys such as the SDSS (see Brosch et al. 2010; Finkelman, Graur \& Brosch 2011). 
Nevertheless, the SDSS holds an enormous potential for detecting many new PRGs at various redshifts among the millions of targeted objects.
However, no automated procedure for searching this database for galaxies with certain characteristics exists at the present. 
To help classify the millions of objects, the interactive project `Galaxy Zoo' invites the public to examine on-line images of galaxies and answer a standard set of questions about their features (Lintott et al.\ 2008).
While there is no unique classification for PRGs, any likely candidate can be posted to `The Possible Polar Ring Thread' or any other related official forum in the Galaxy Zoo (http://www.galaxyzoo.org/). 

Galaxy Zoo therefore provides an opportunity to significantly increase our knowledge of nearby PRGs.
Moiseev et al.\ (2011) selected from the dedicated web forums a `reference' galaxy sample which was used to formulate criteria for selecting new nearby candidate PRGs from the Galaxy Zoo. 
After visually inspecting thousand of objects, Moiseev et al.\ compiled a final list of 275 candidate PRGs and related objects (the Sloan-based Polar Ring Catalogue, SPRC), including 14 galaxies provided by us.
The new candidates are typically fainter and are more distant than the previously known PRGs.
Their study, therefore, holds a promise of building up a more complete picture of how these systems form.

This paper is organized as follows: Section \ref{S:sample} explains the sample selection, Section \ref{S:Obs_and_Red}
gives a description of all the observations and data analysis, and Section \ref{S:analysis} details the properties of individual galaxies. We discuss the results in Section \ref{S:discuss}.

\section{Sample}
\label{S:sample}
The sample, included also in the SPRC, was selected by us from a large compilation of suspected PRG publicly available on `The Possible Polar Ring Thread' web forum. 
We carefully selected 16 promising PRGs by visually
examining their SDSS images. These candidates were chosen following the confident identification
of an extended polar structure (about 10-20 arcsec wide). 
We also verified that the apparent polar ring in each
sample galaxy is not produced by a projected background galaxy by using their SDSS spectra. 
The sample galaxies are listed in Table 1 with their coordinates, radial velocity and $g$-magnitude taken from the SDSS database.

\begin{table*}
 \centering
  \caption{Global parameters for galaxies in our sample.
  \label{t:Obs}}
\begin{tabular}{rlllrll}
\hline
PRG \# &  \multicolumn{1}{c}{Object}     & RA       & DEC    & v$_{Helio}$  & $m_g$ & Type of nucleus \\
{}     & {}       & (J2000.0)& (J2000.0) & (km/s)     &   (mag)     \\
\hline
{1} & SDSS J003209.80+010836.5 & 00h32m09.8s  & +01d08m37s &   17715  & 16.7 & SF\\
{2} & SDSS J004812.18-001255.5 & 00h48m12.2s  & -00d12m56s &   16925  & 17.1 & Composite\\
{3} & SDSS J024258.42-005709.3 & 02h42m58.4s  & -00d57m09s &   12888  & 17.8 & SF \\
{4} & {SDSS J082038.18+153659.7} & {08h20m38.2s}  & {+15d37m00s} & 12736 &   {17.2} & Low S/N LINER \\
{5} & {SDSS J084832.00+322012.4} & {08h48m32.0s}  & {+32d20m12s} & 19734 & {16.8} & Low S/N LINER \\
{6} & {SDSS J091453.64+493824.0} & {09h14m53.6s}  & {+49d38m24s}  & 9522 & {16.2} & Low S/N SF \\
{7} & {SDSS J094207.34+362417.2} & {09h42m07.3s}  & {+36d24m17s}  & 18029 & {17.9} & Low S/N SF\\
{8} & {SDSS J094302.33-004850.0} & {09h43m02.3s}  & {-00d48m50s}  & 20270 & {16.9} & AGN \\
{9} & {SDSS J104623.65+063710.2} & {10h46m23.6s}  & {+06d37m10s}  & 8331 & {16.9}  & SF \\
{10} & {SDSS J114444.02+230944.9} & {11h44m44.0s}  & {+23d09m45s}  & 14509 & {17.1} & Low S/N LINER \\
{11}& {SDSS J115228.29+050044.7} & {11h52m28.3s}  & {+05d00m45s}  & 23189 & {17.7} & Composite\\
{12}&{SDSS J130816.92+452235.1} & {13h08m16.9s}  & {+45d22m35s}  & 8792 & {17.1} & Low S/N SF \\
{13}&{SDSS J135941.70+250046.0} & {13h59m41.7s}  & {+25d00m46s}  & 9370 & {15.7} & Low S/N LINER \\
{14}&{SDSS J151114.08+370237.6} & {15h11m14.1s}  & {+37d02m38s}  & 16499 & {15.7} & AGN \\
{15}& SDSS J204805.66+000407.8  & 20h48m05.6s & +00d04m08s   &   7396 & 16.2 & AGN\\
{16}& SDSS J212339.15-002235.2 & 21h23m39.1s & -00d22m35s & 18582 & 17.8 & SF \\
\hline
\end{tabular}
\end{table*}

\section{Observations and data analysis}
\label{S:Obs_and_Red}
Observations of 11 galaxies were obtained in October 2010 and March 2011 with the 1.8-m Vatican Advanced
Technology Telescope (VATT) on Mount Graham. A 4K CCD camera was used with a pixel scale of 0.37 arcsec per pixel, yielding a field of view of $12 \times 12$ arcmin$^2$. The observations were performed using the Sloan $u$- and $r$-band filters with typical exposure times of 20 and 10 min, respectively.
Five of our sample galaxies lie close to the celestial equator in the Southern Galactic Cap and were therefore included in the deep 275 deg$^2$ Stripe82 subsample of the SDSS. The multiple photometric scans of this part of the sky can be co-added to reach imaging roughly $\sim$2 mag deeper than single SDSS scans.

Image reduction was performed with standard tasks within {\small IRAF}\footnote{IRAF is distributed by the National Optical Astronomy Observatories (NOAO), which is operated by the Association of Universities, Inc.\ (AURA) under co-operative agreement with the National Science Foundation}. These include bias subtraction, overscan subtraction and flatfield correction. 
SDSS images were reduced and flux-calibrated using the standard pipelines. 
By measuring the flux of foreground stars in the fields of single SDSS frames we were able to calibrate the deep Stripe82 and VATT images.
Contour maps of the $r$-band images of our sample galaxies are presented in Fig.\ 1.

Using our narrow-band H$\alpha$ filter set we were able to cover the rest-frame H$\alpha$+[NII] emission of seven of our sample galaxies.
The H$\alpha$ images include photons from the H$\alpha$ and the [NII] lines and from the continuum. 
To obtain a continuum free H$\alpha$ image for each galaxy the $r$-band image was scaled to match the intensity of the stellar continuum in the narrow-band image and was subtracted from the narrow-band image. 
The continuum-subtracted images are shown in Fig.\ 2 and reveal extended emission along the polar structure in four of the galaxies.
About 2 arcmin north of PRG 15 we also detect H$\alpha$ emission from a disrupted galaxy (SDSS J204805.48+000702.5). 

Unfortunately, the small angular size of the objects, combined with the angular resolution of our images, do not allow a detailed morphological analysis of our sample galaxies.
Since we cannot simply separate the light of the luminous stellar ring and the host, we attempt only a crude decomposition of the two distinct components. 
We first draw isophotal contours of each galaxy in single SDSS images and visually fit an elliptical contour for the $\mu_g =25$ mag arcsec$^{−2}$ isophote of the hosts (see fig.\ 5 in Finkelman, Graur \& Brosch 2011). The integrated magnitudes of the hosts are then computed within the area delimited by the these elliptical contours.
We restrict the ring photometry to the parts that are not projected against the bright regions of the host galaxies.
This is done by computing the integrated light enclosed within the area where the surrounding ring is brighter than $\mu_g =25$ mag arcsec$^{−2}$ and subtracting the measured light from the host.
For the cases of short polar rings, which are almost hidden by the light of the central host, we set the delimiting isophotal contour at $\mu_g =24$ mag arcsec$^{−2}$. 
The integrated magnitudes, the $u$-$r$ colours and the major-axis position angles (PAs, measured from north through east) of the host and ring of each galaxy are listed in Table 2.
Our measurements show that the central hosts are brighter by $\sim$2-3 mag than the rings, although the true differences are probably smaller since the hosts are contaminated by light from the rings. 
The colours of the rings span a $u-r$ range from 1.40 to 2.42 and are, on average, bluer by $0.60\pm0.27$ than the colour of the hosts which vary from 2.03 to 2.93.

\begin{figure*}
\begin{center}
\begin{tabular}{ccc}
 \includegraphics[width=6cm]{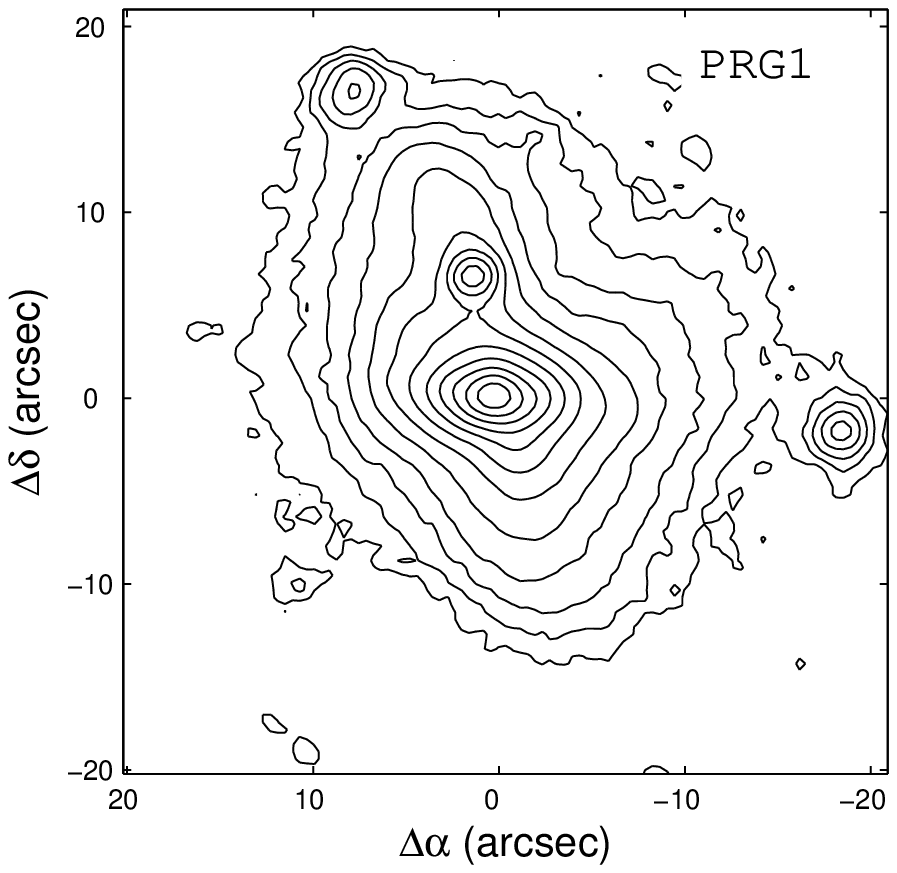} &  \includegraphics[width=6cm]{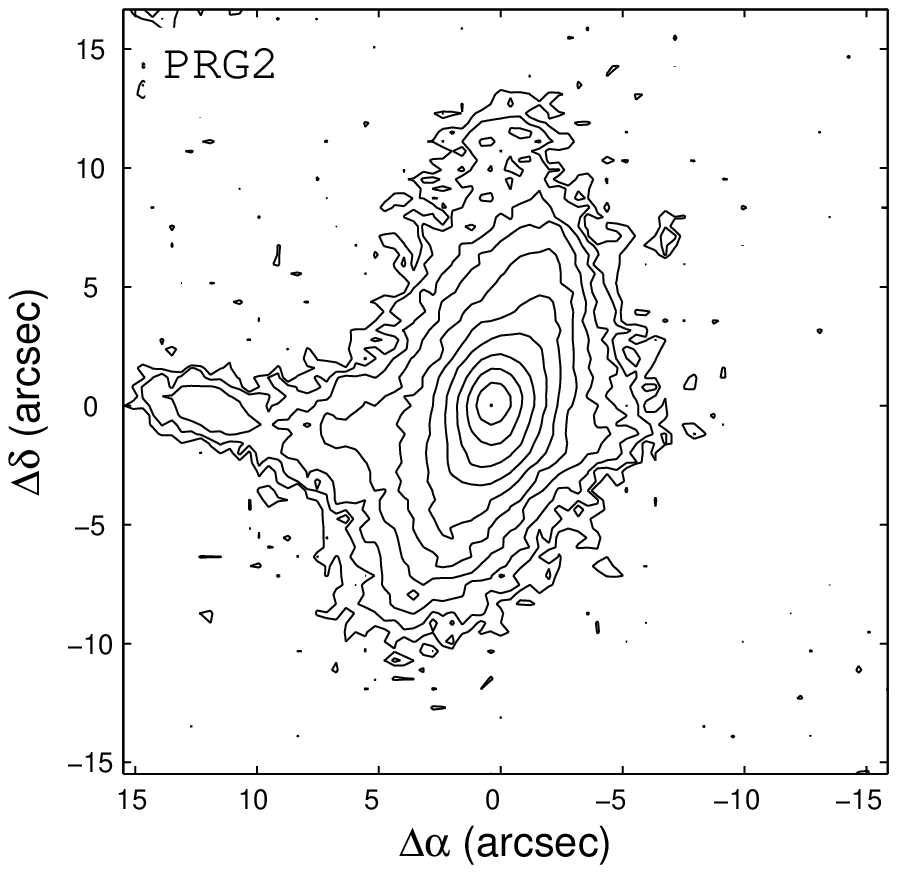} &\includegraphics[width=6cm]{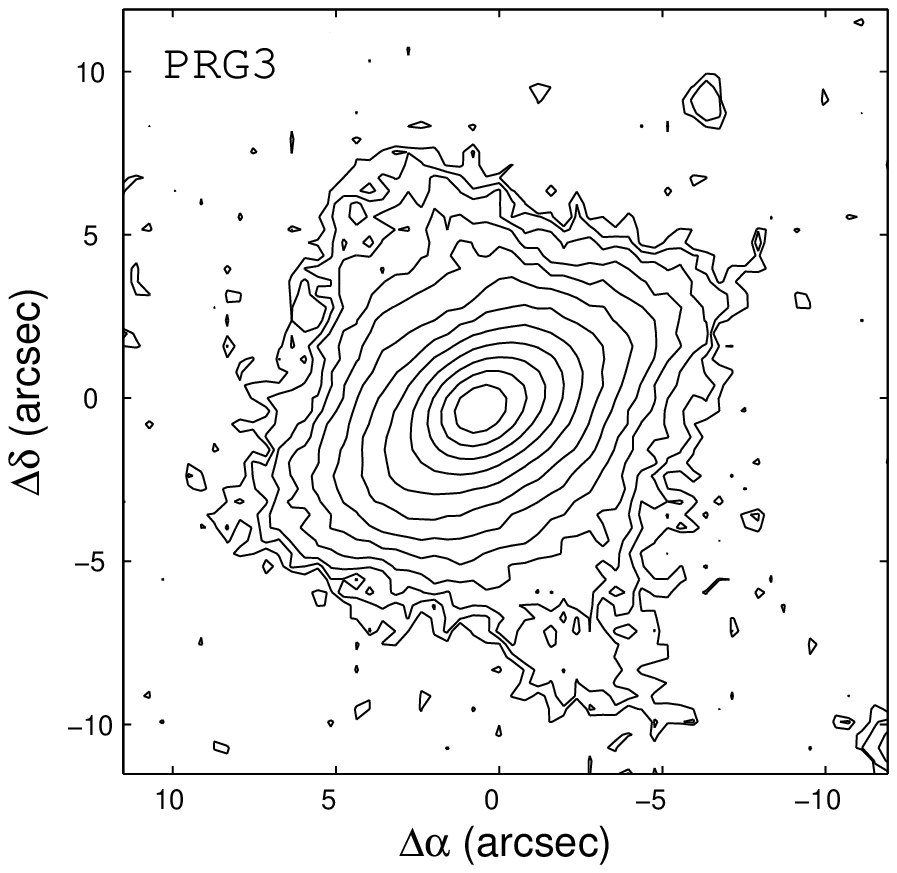}\\
  \vspace{-3mm}
 \includegraphics[width=6cm]{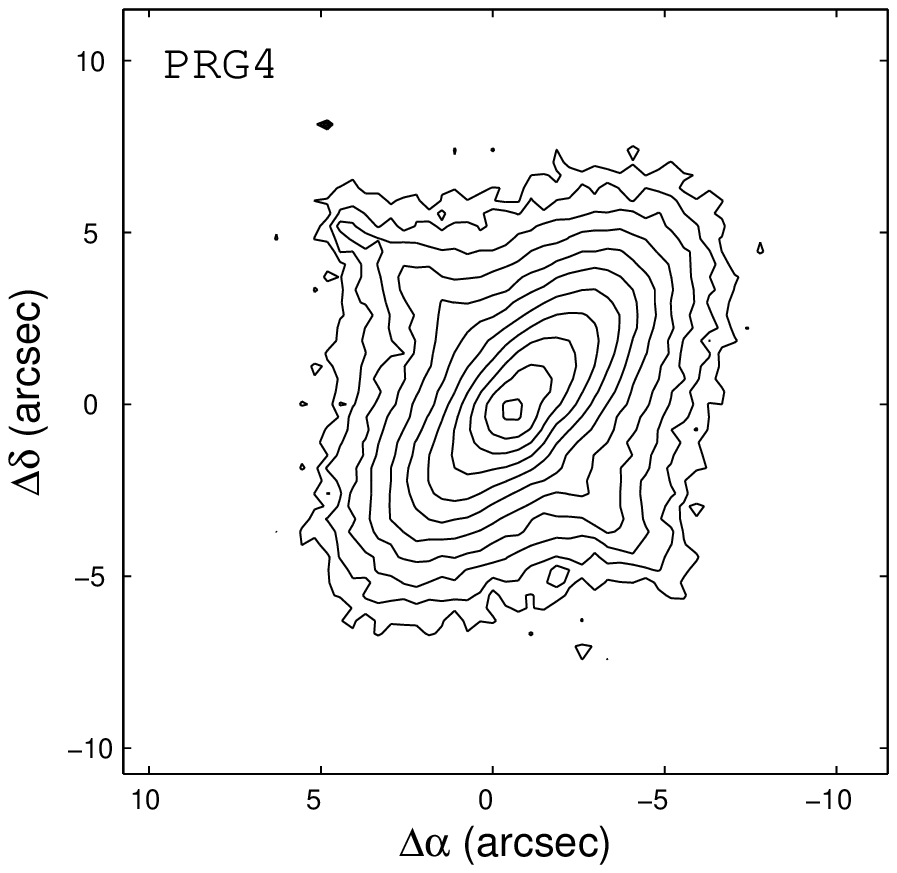} & \includegraphics[width=6cm]{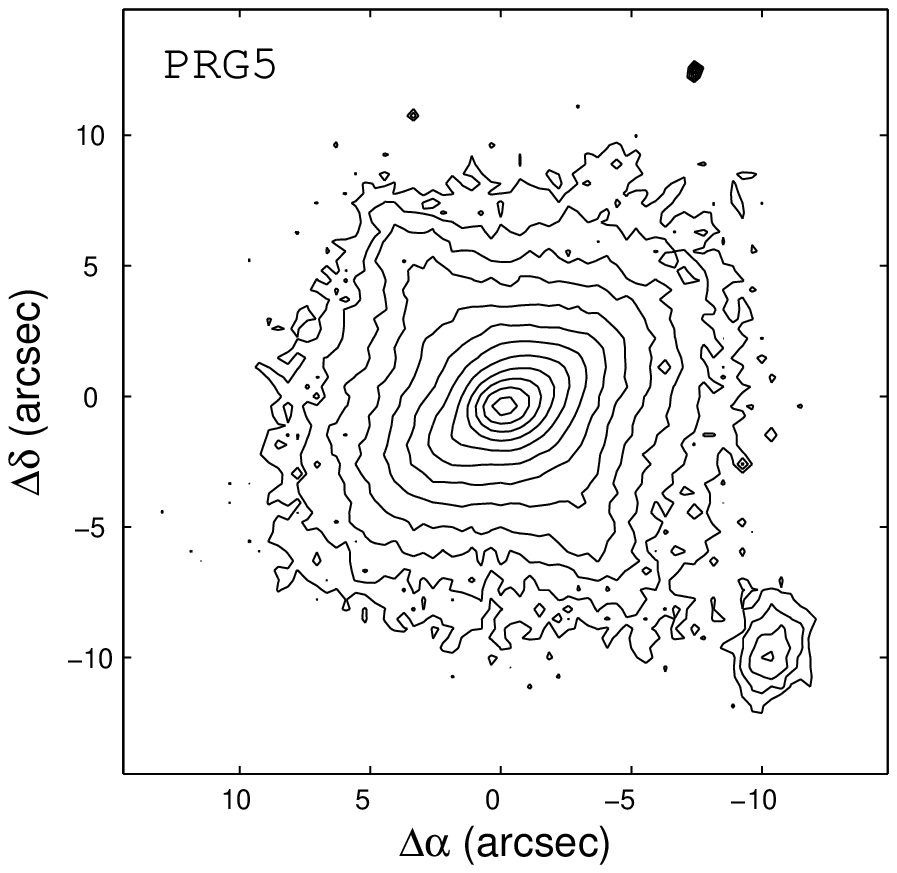} & \includegraphics[width=6cm]{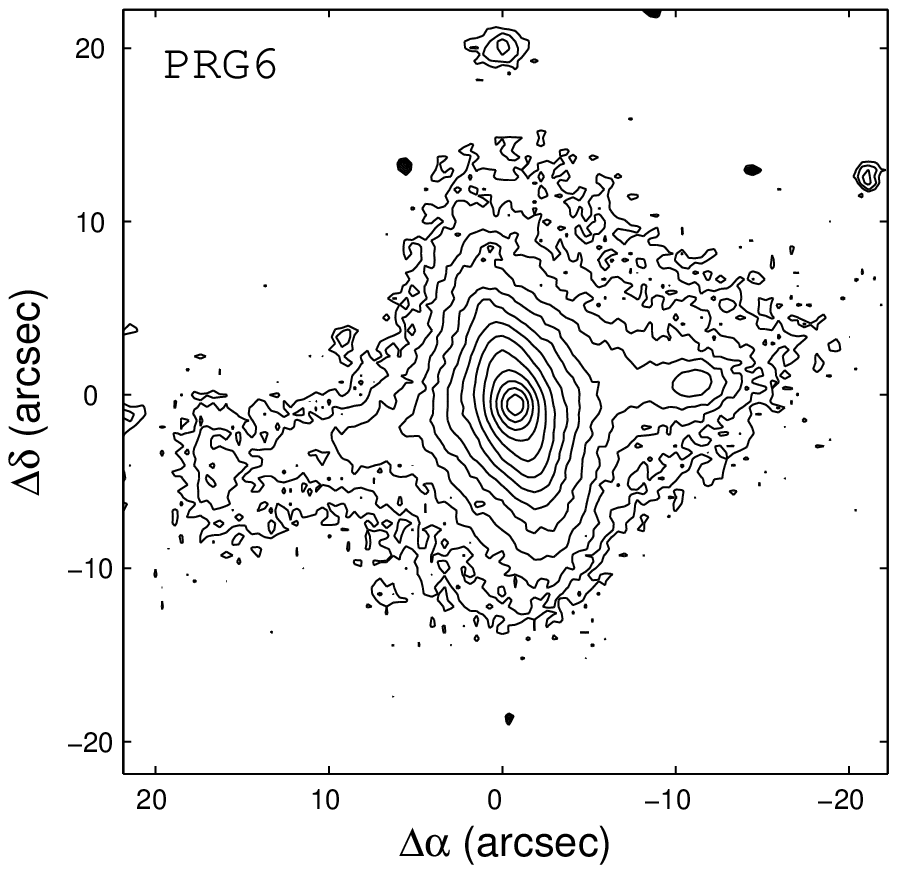}  \\
  \vspace{-3mm}
  \includegraphics[width=6cm]{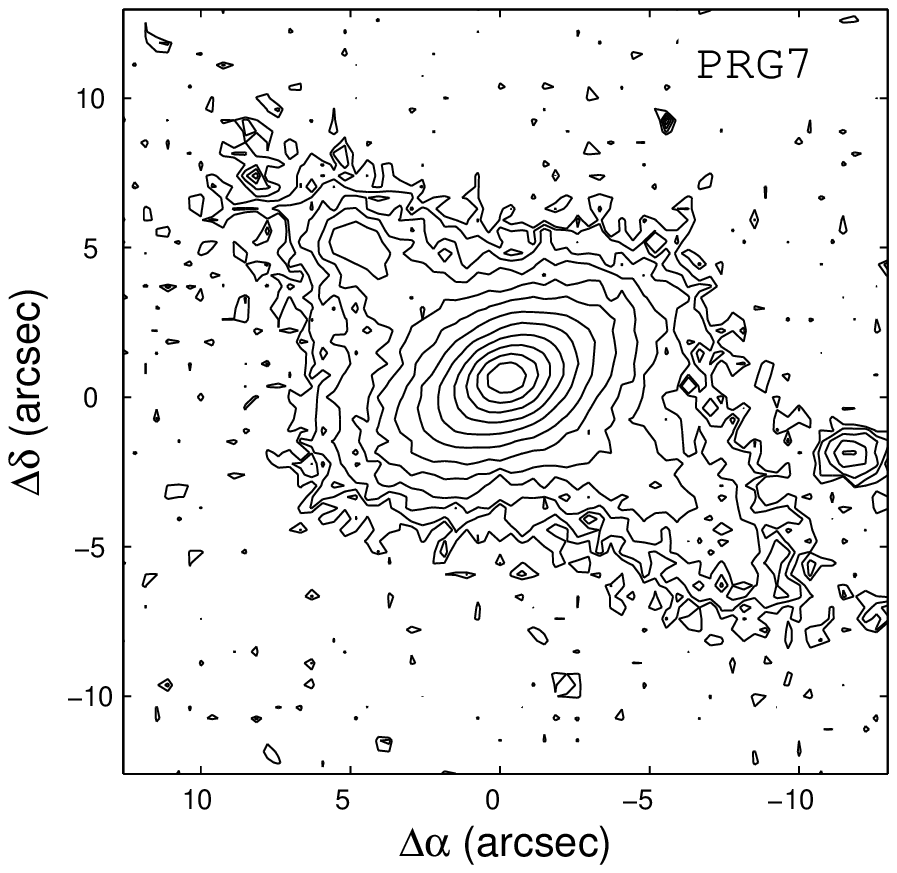} & \includegraphics[width=6cm]{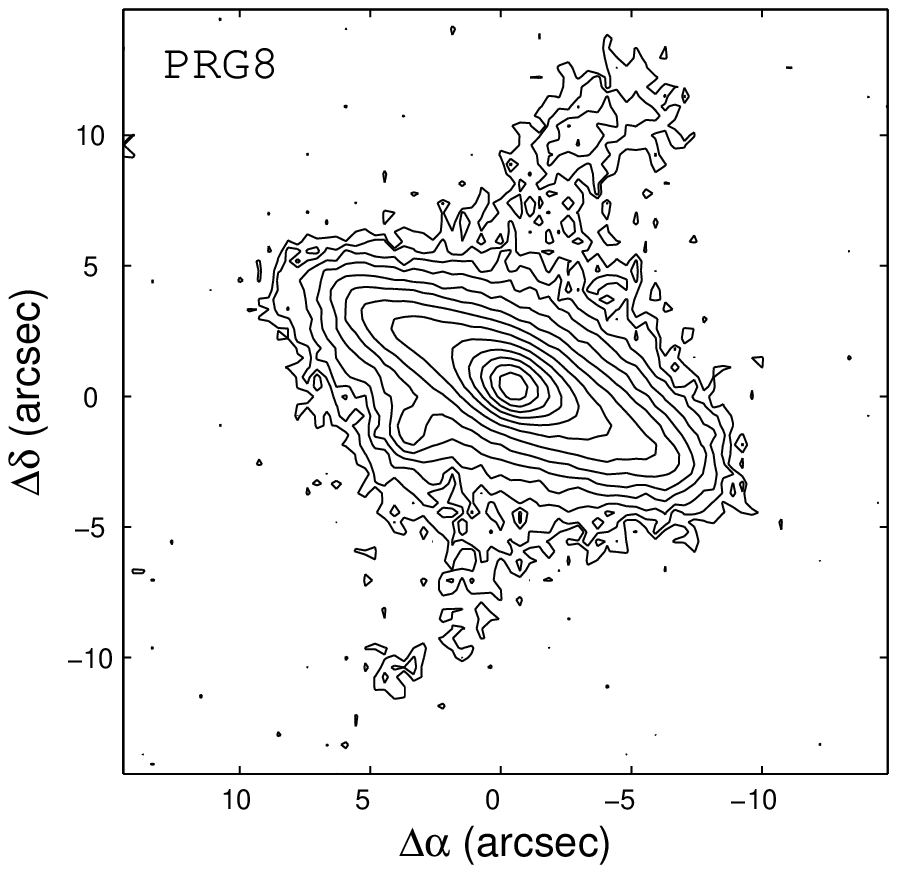} & \includegraphics[width=6cm]{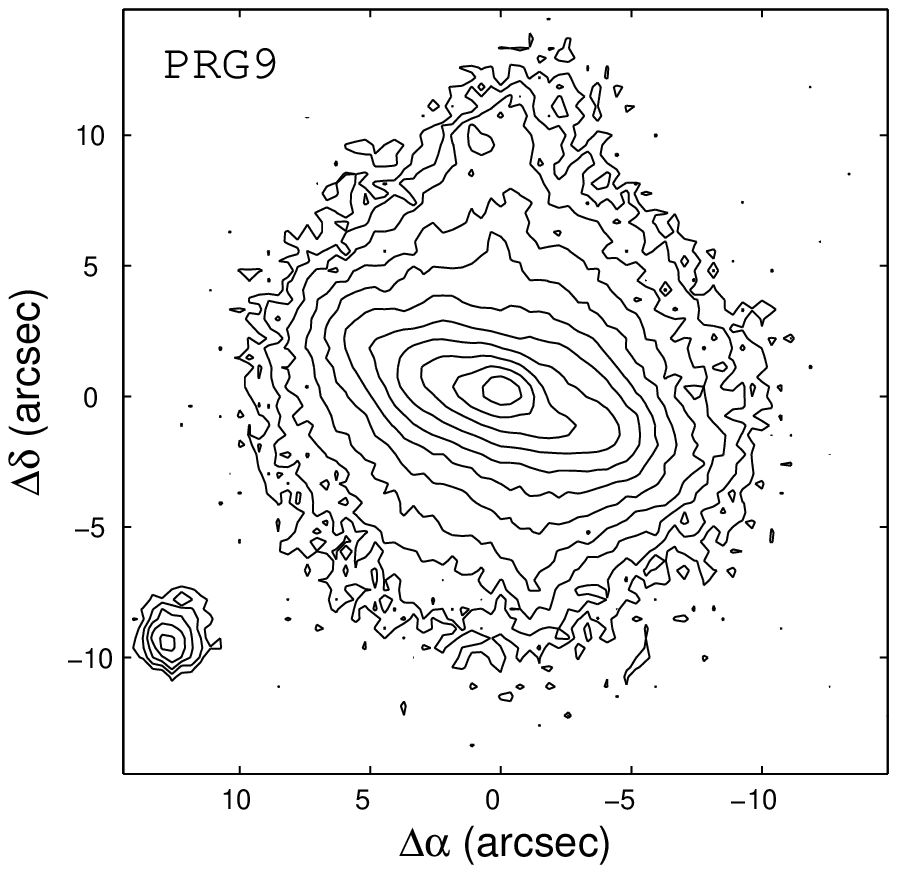}
\end{tabular}
\end{center}
\caption{Contour maps of the $r$-band images of our sample galaxies. }
 \label{f:Rmaps}
\end{figure*}
\setcounter{figure}{0}
\begin{figure*}
\begin{center}
\begin{tabular}{ccc}
\includegraphics[width=6cm]{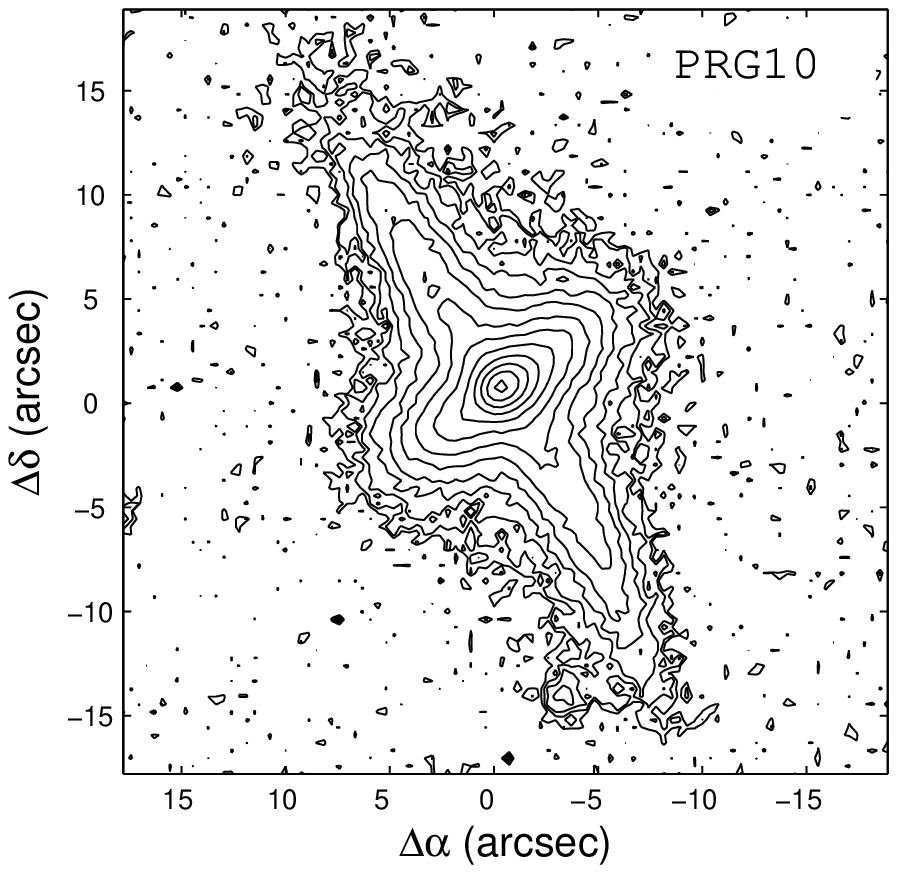} & \includegraphics[width=6cm]{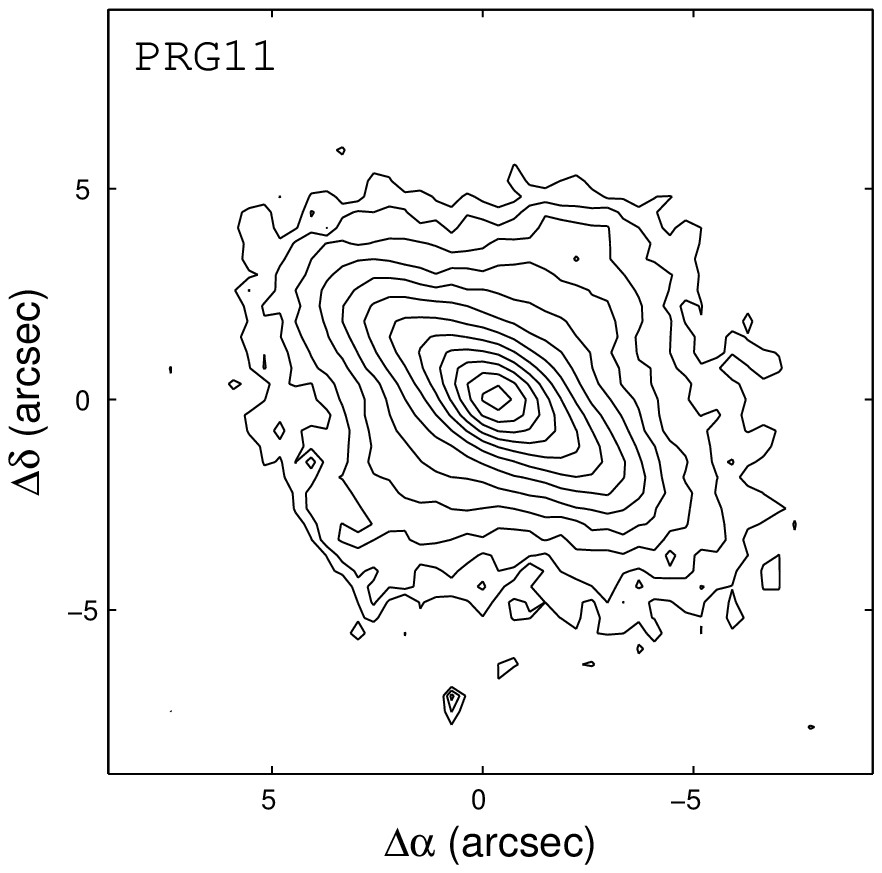} & \includegraphics[width=6cm]{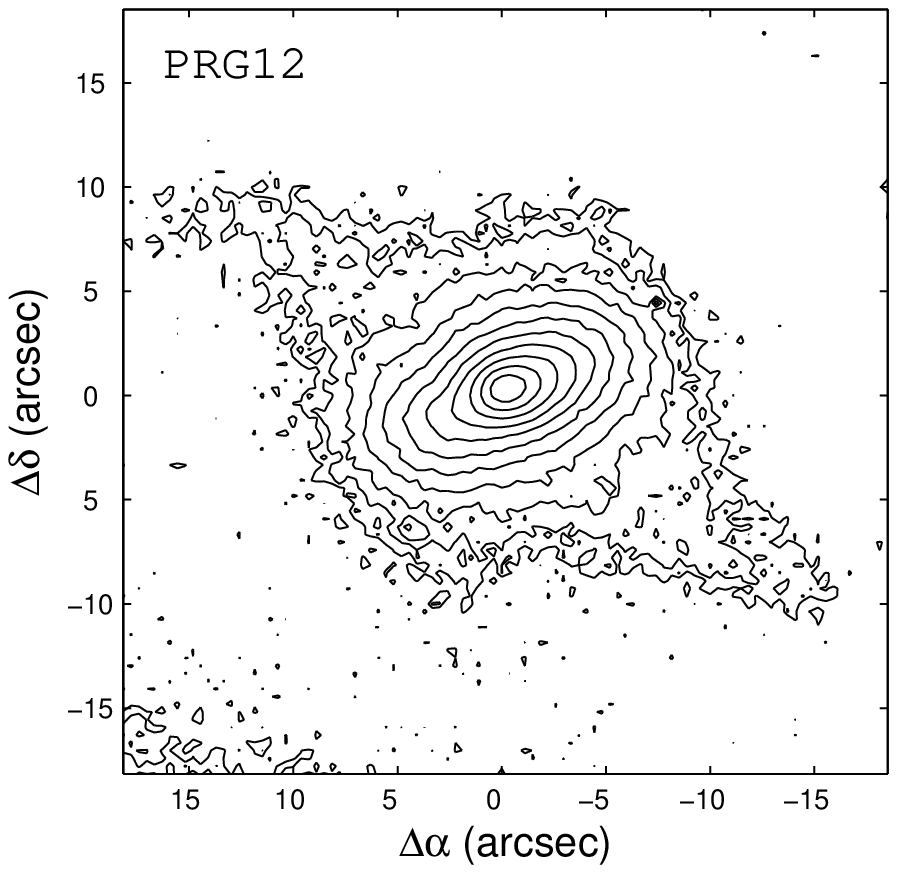}\\
  \vspace{-3mm}
\includegraphics[width=6cm]{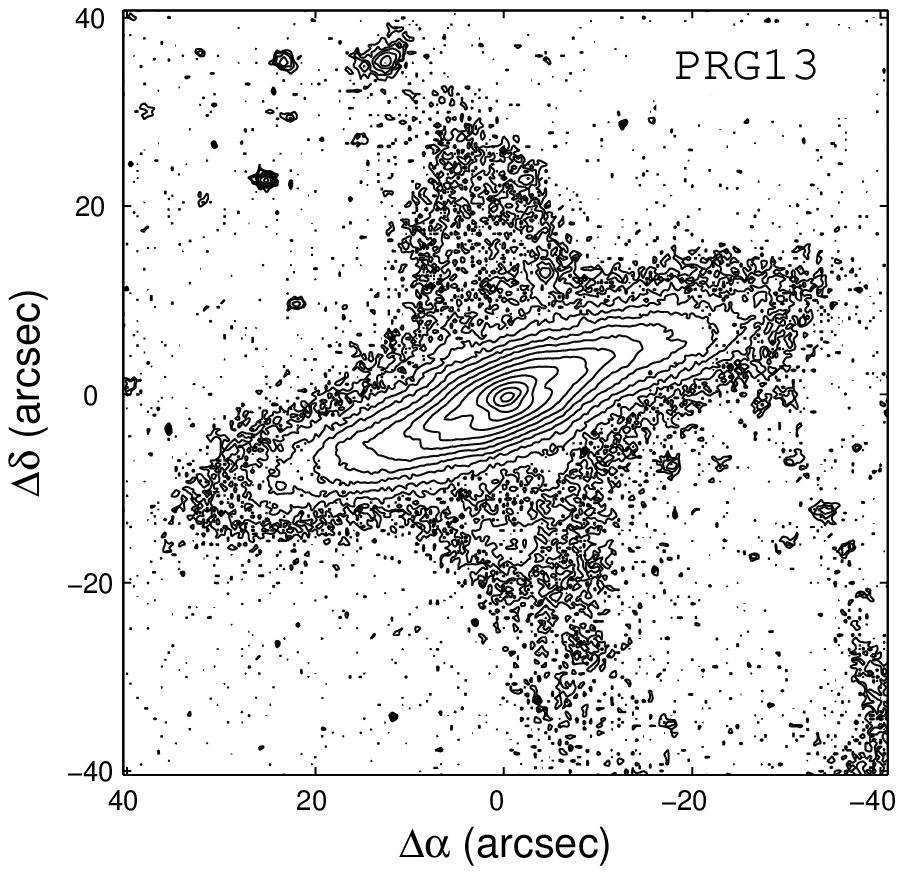} & \includegraphics[width=6cm]{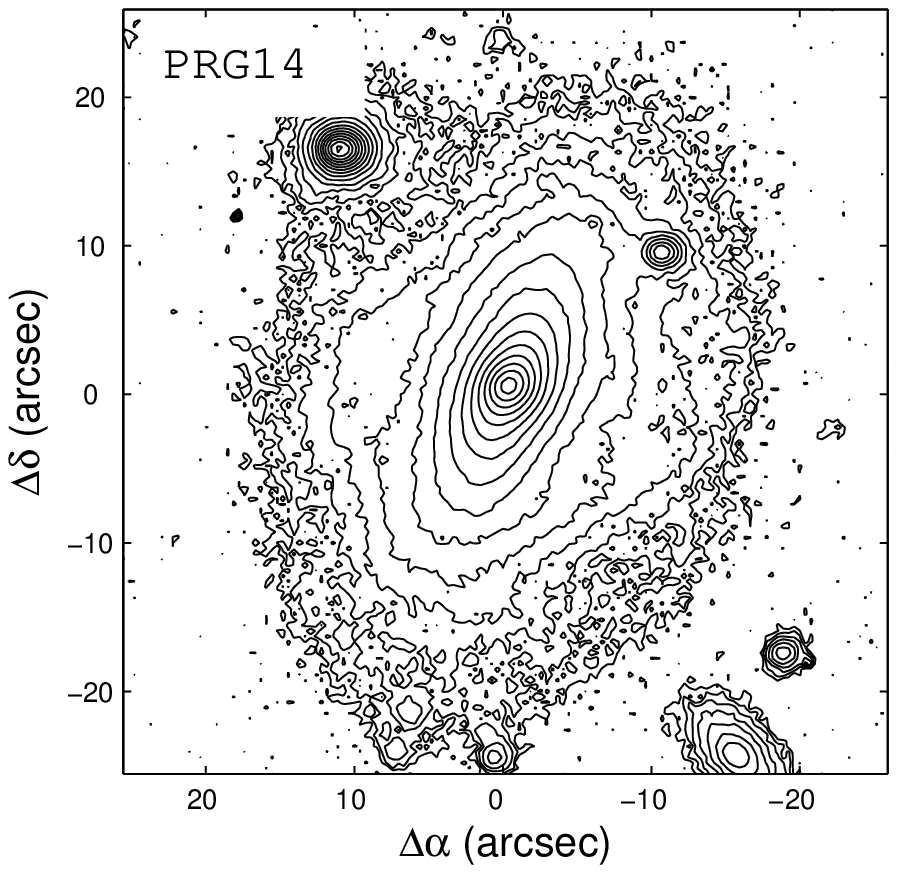} & \includegraphics[width=6cm]{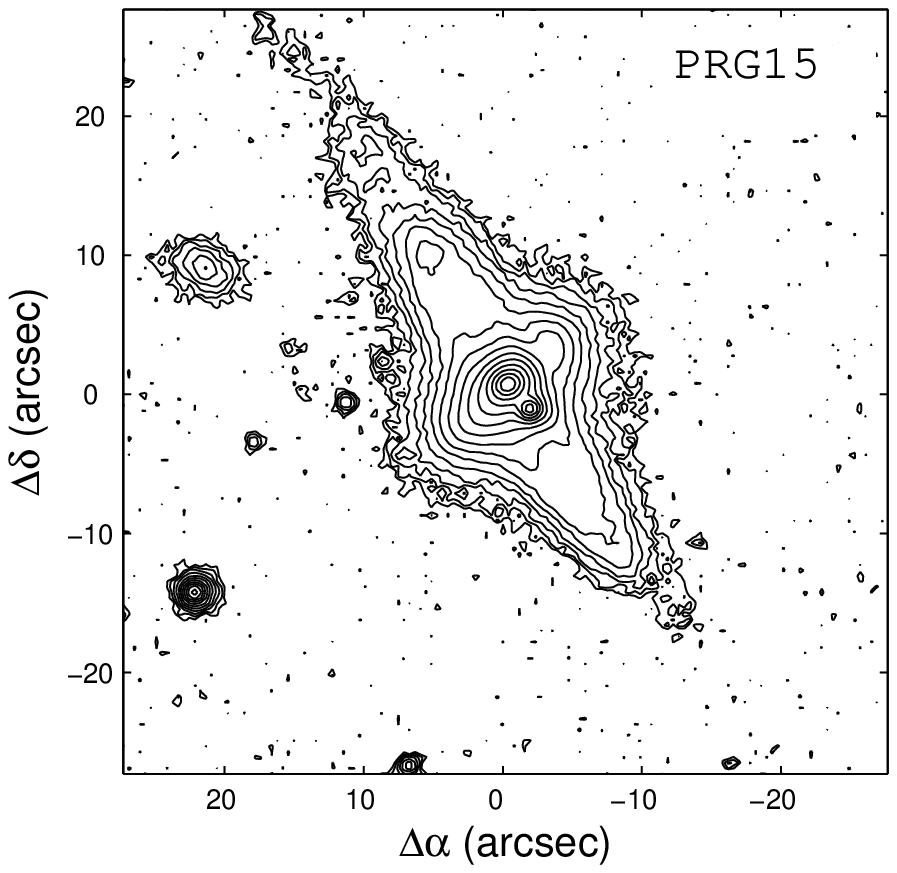} \\
  \vspace{-3mm}
\includegraphics[width=6cm]{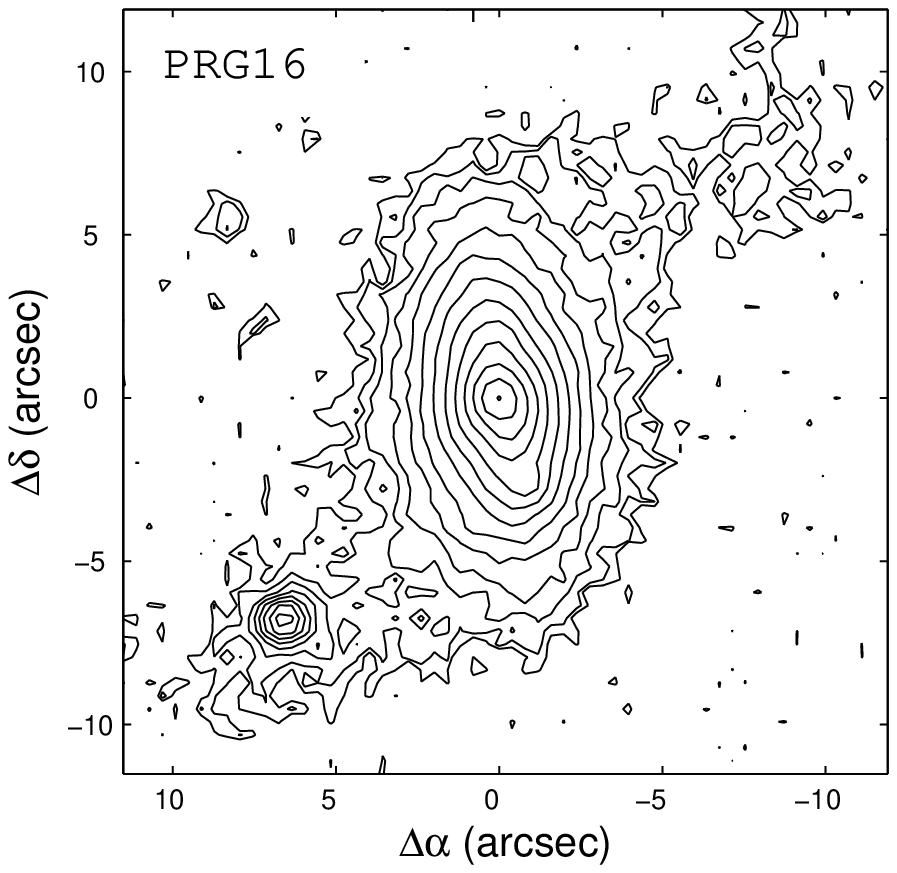} & {} & {} 
\end{tabular}
\end{center}
\caption{cont.}
 \label{f:}
\end{figure*}

\begin{figure*}
\begin{center}
\begin{tabular}{ccc}
\includegraphics[width=6cm]{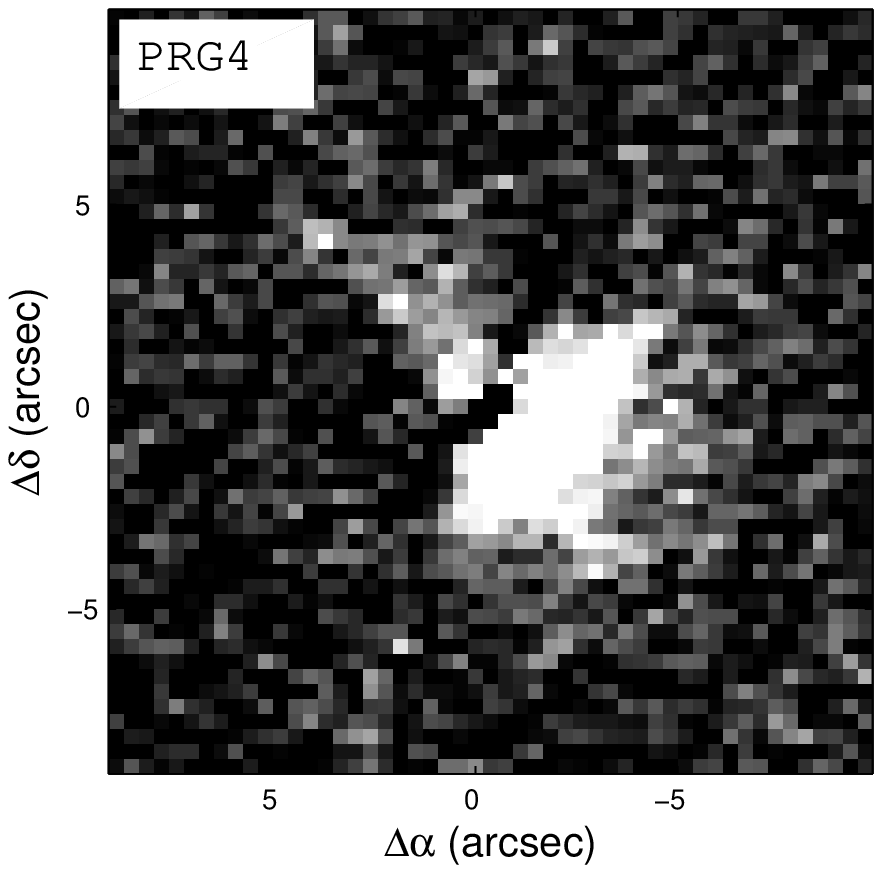} & \includegraphics[width=6cm]{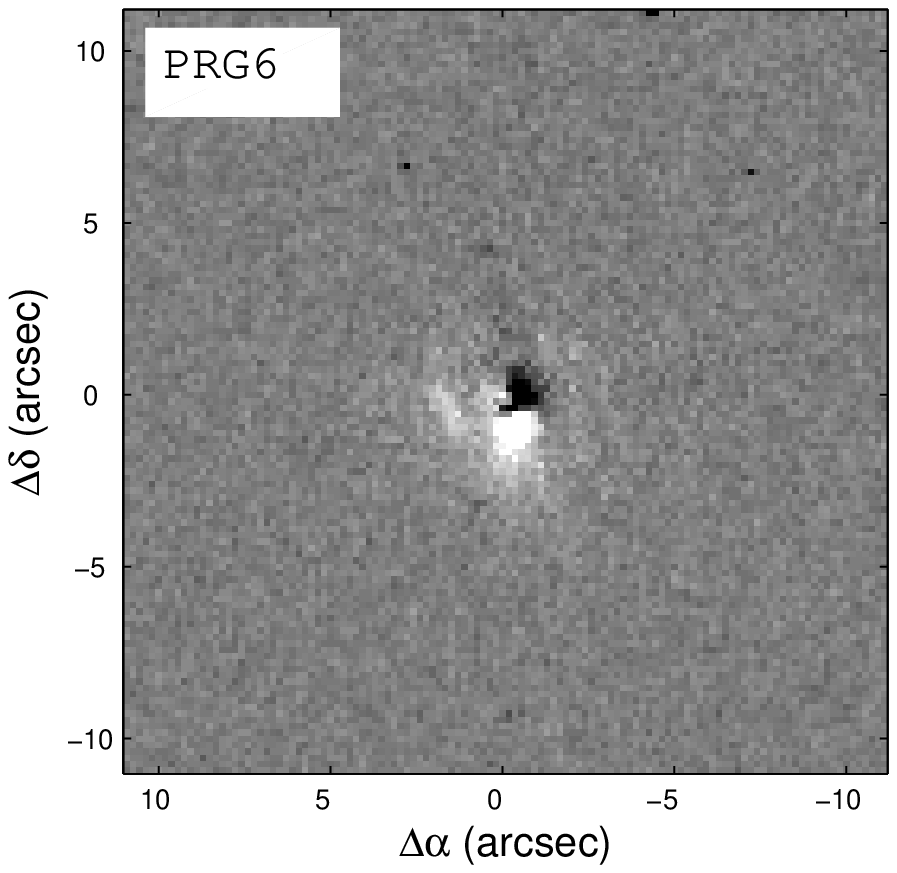} & \includegraphics[width=6cm]{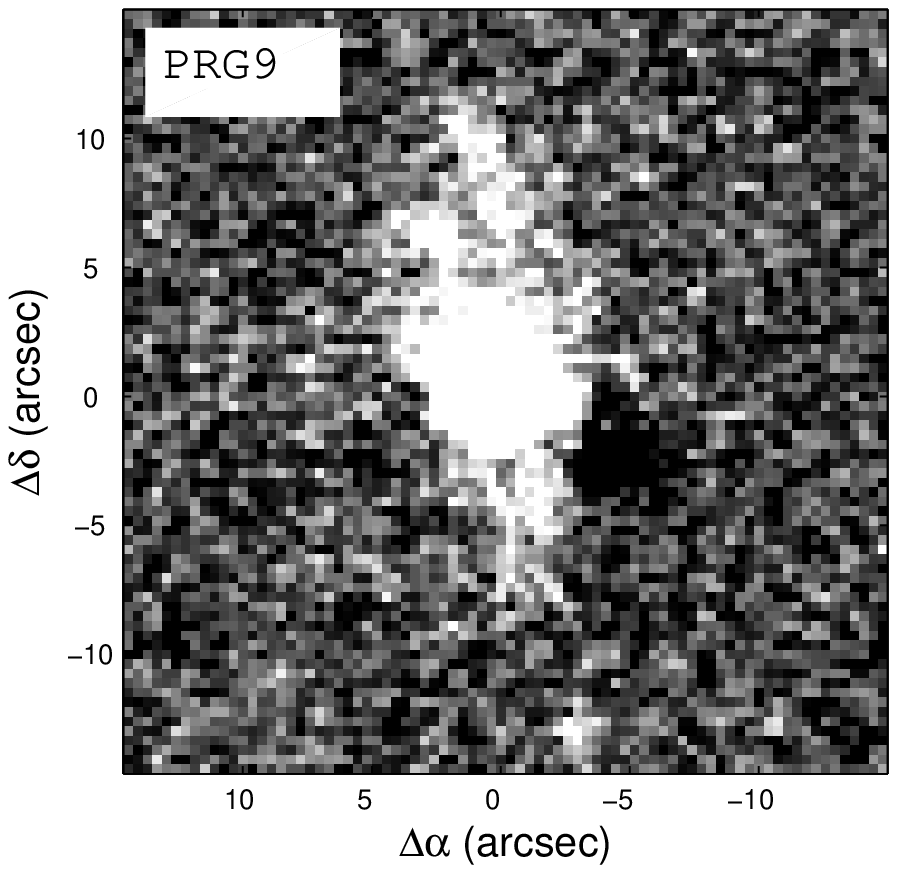} \\
 \vspace{-3mm}
\includegraphics[width=6cm]{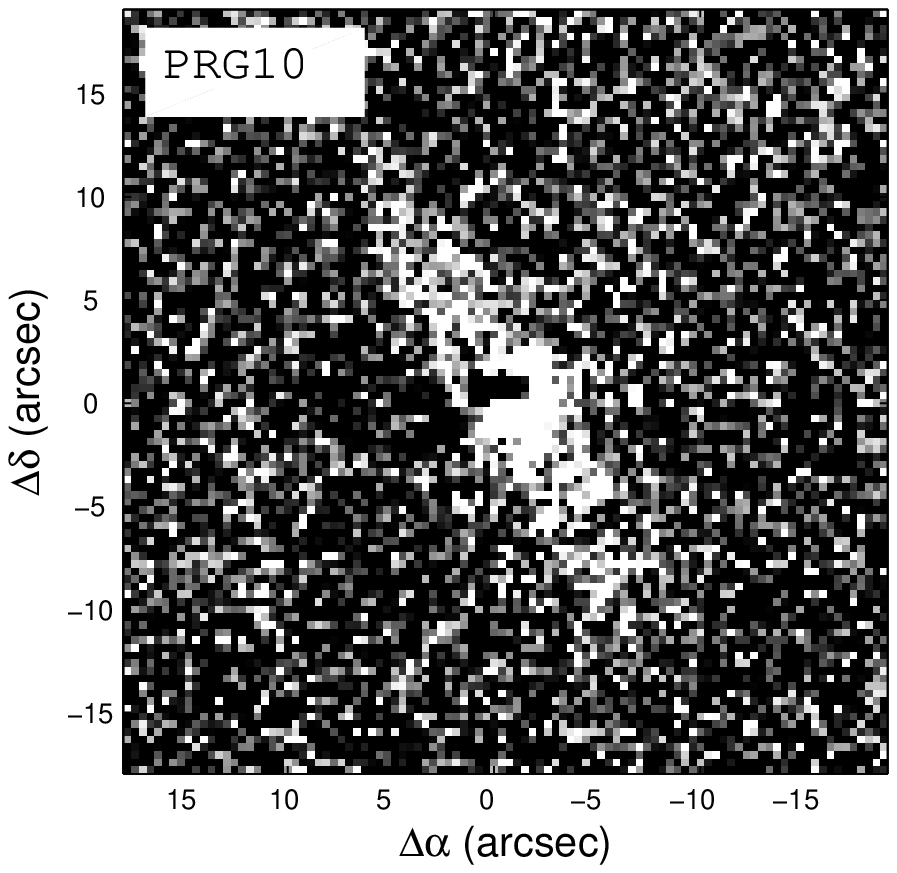} & \includegraphics[width=6cm]{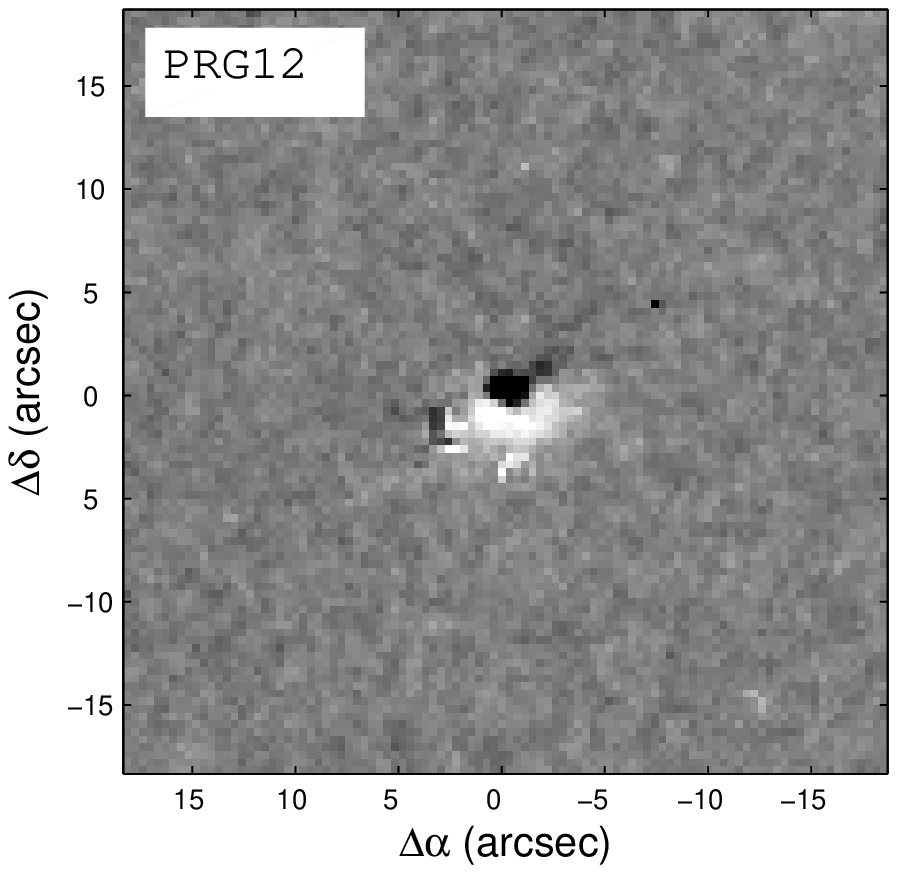} & \includegraphics[width=6cm]{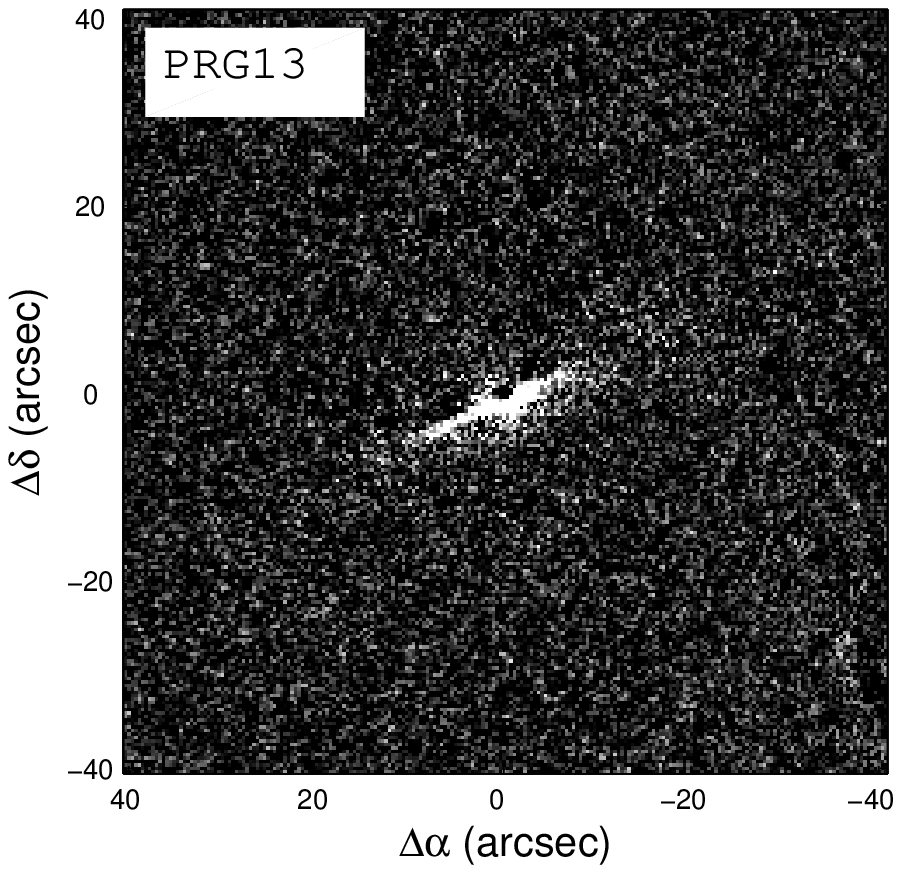}\\ 
 \vspace{-3mm}
\includegraphics[width=6cm]{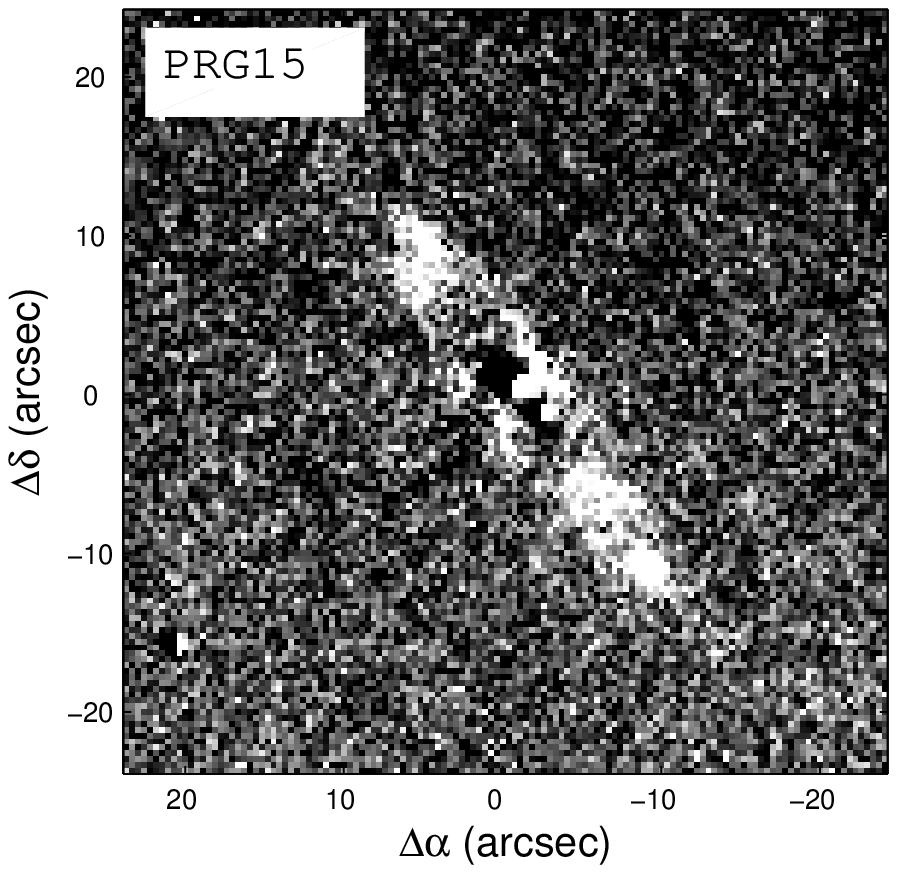} & {} & {}
\end{tabular}
\end{center}
\caption{Continuum-subtracted H$\alpha$+[NII] images. 
Note that the inner $\sim3$ arcsec region in each image is badly subtracted.}
 \label{f:Rmaps}
\end{figure*}

\begin{figure*}
\begin{center}
\begin{tabular}{ccc}
 \includegraphics[width=6cm]{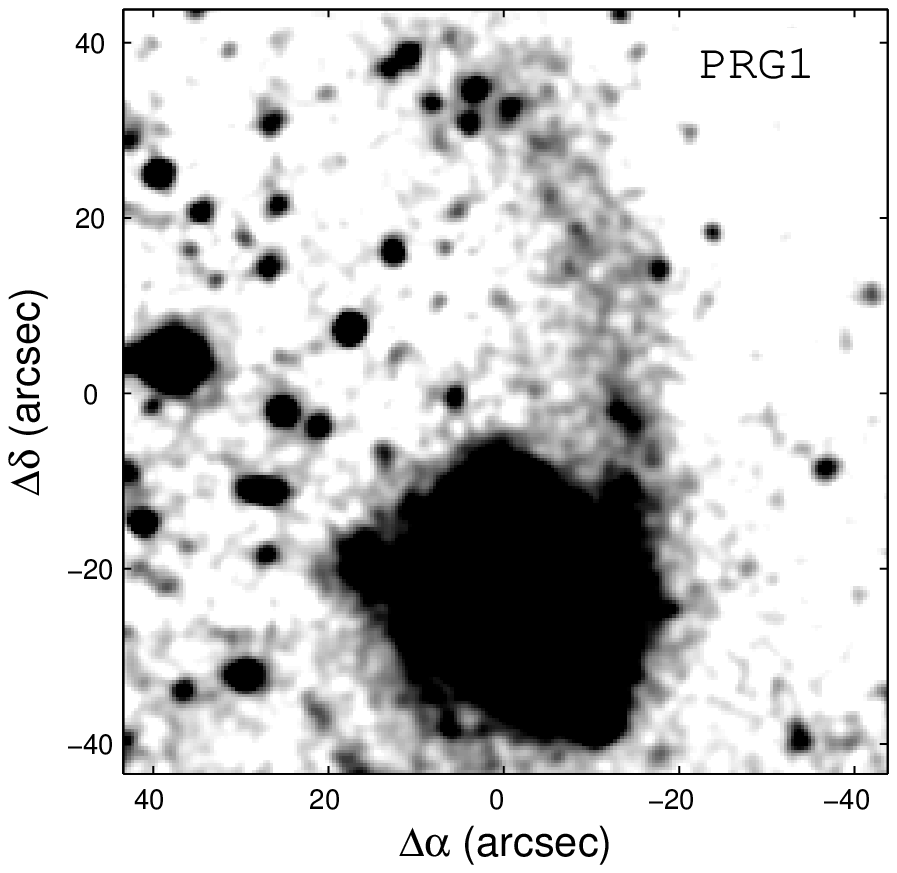} &  \includegraphics[width=6cm]{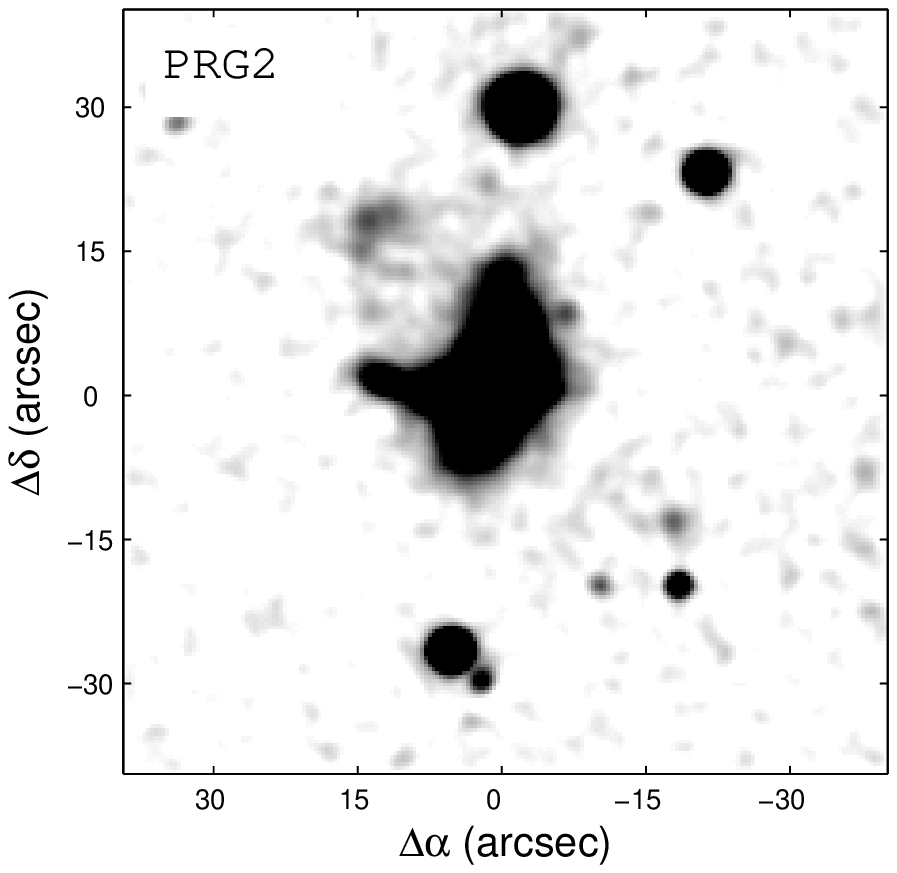} &\includegraphics[width=6cm]{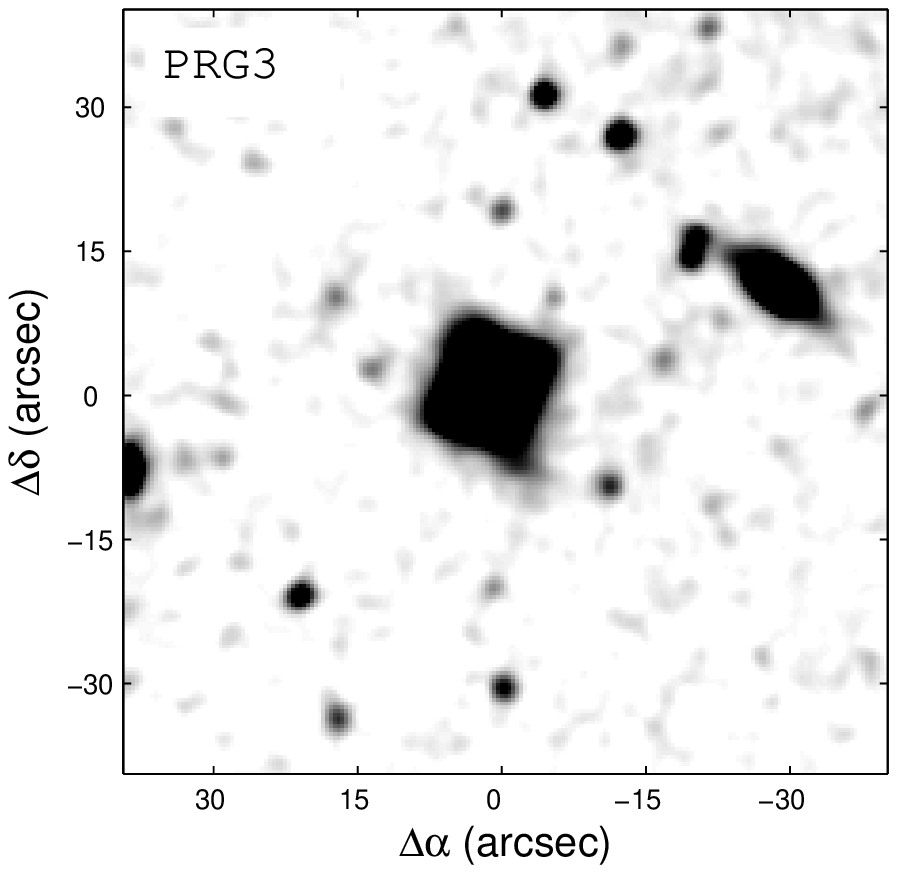}\\
  \vspace{-3mm}
 \includegraphics[width=6cm]{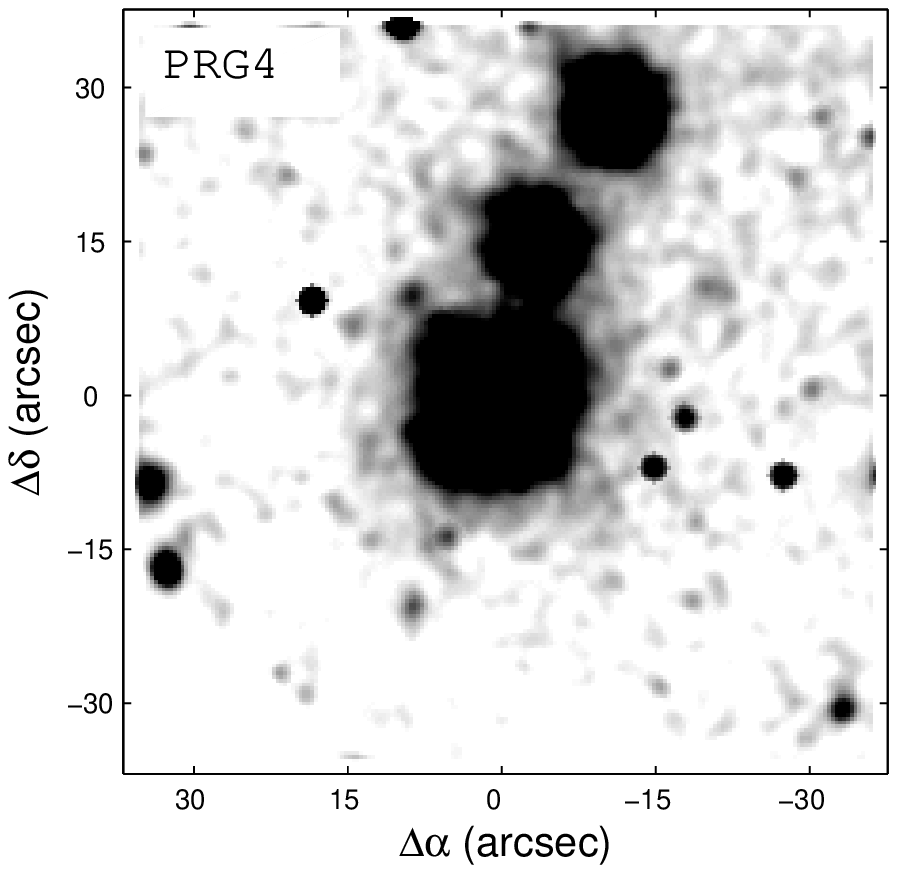} & \includegraphics[width=6cm]{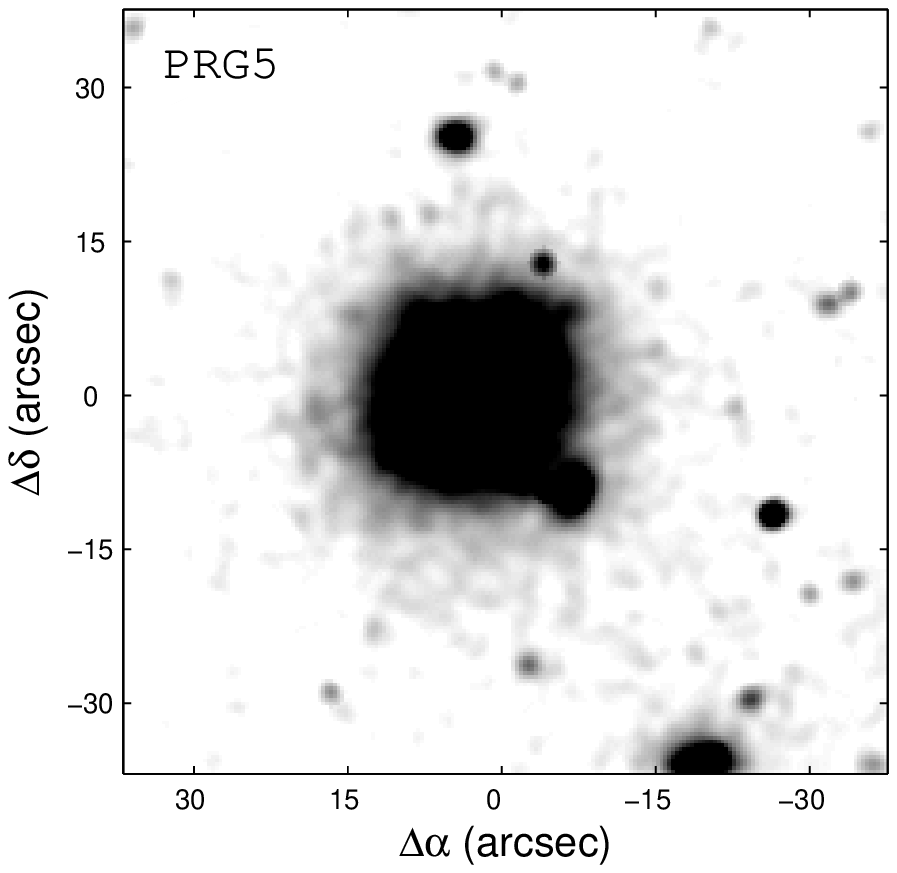} & \includegraphics[width=6cm]{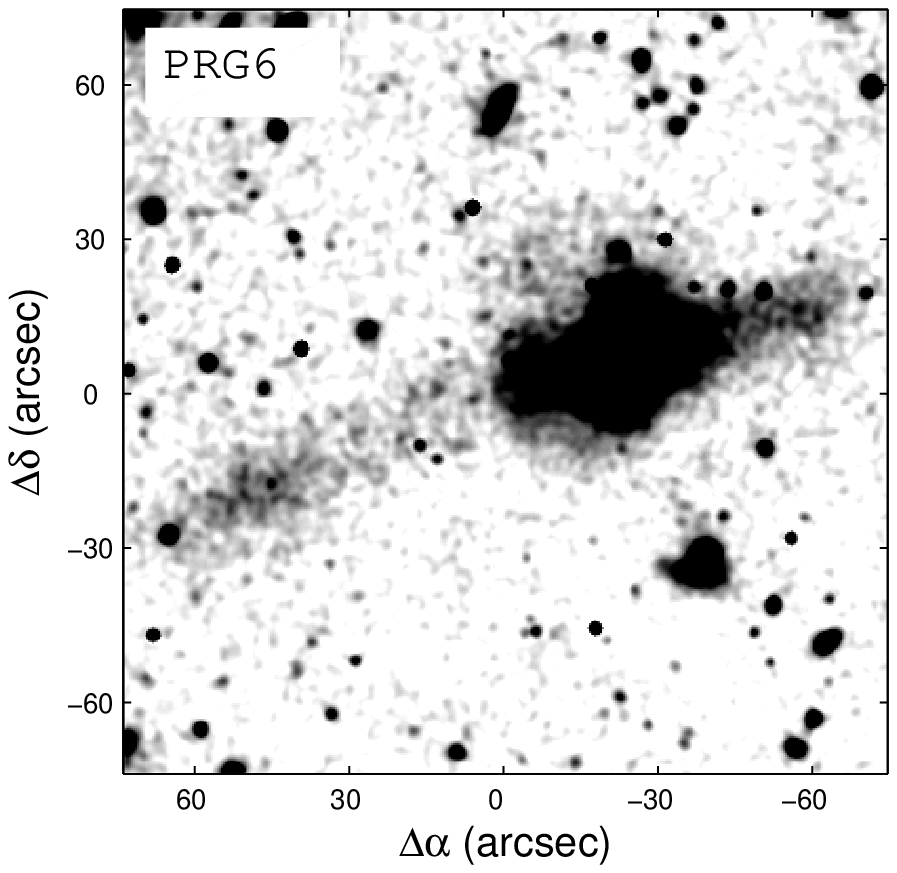} \\
  \vspace{-3mm}
  \includegraphics[width=6cm]{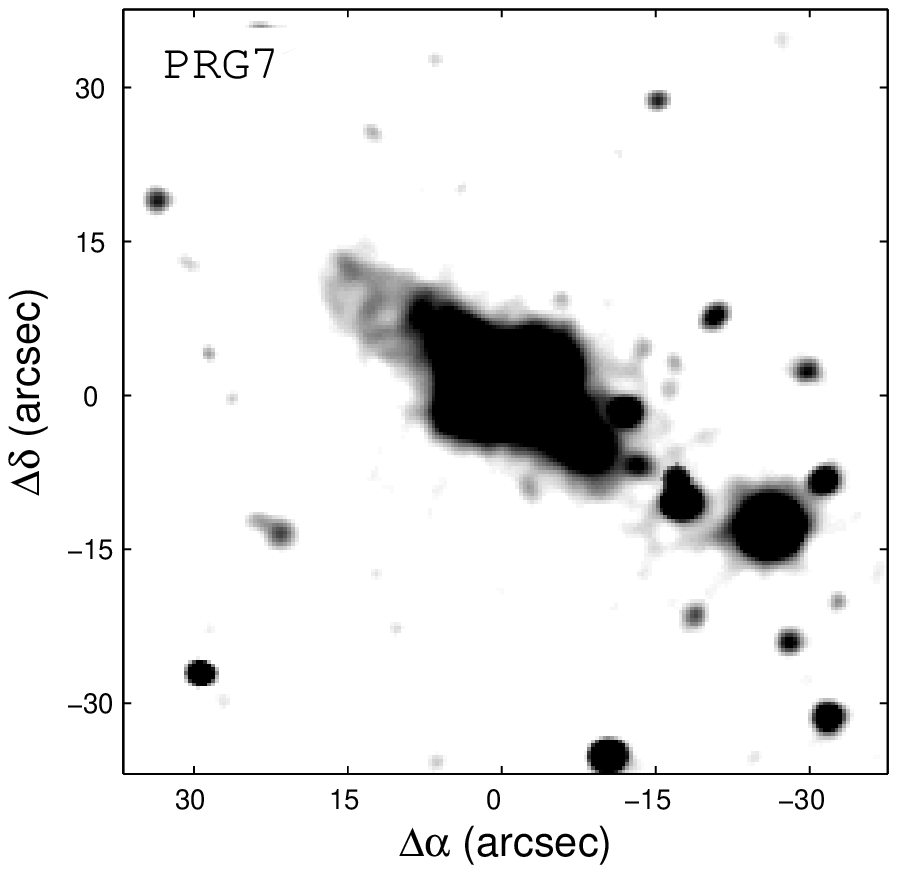} & \includegraphics[width=6cm]{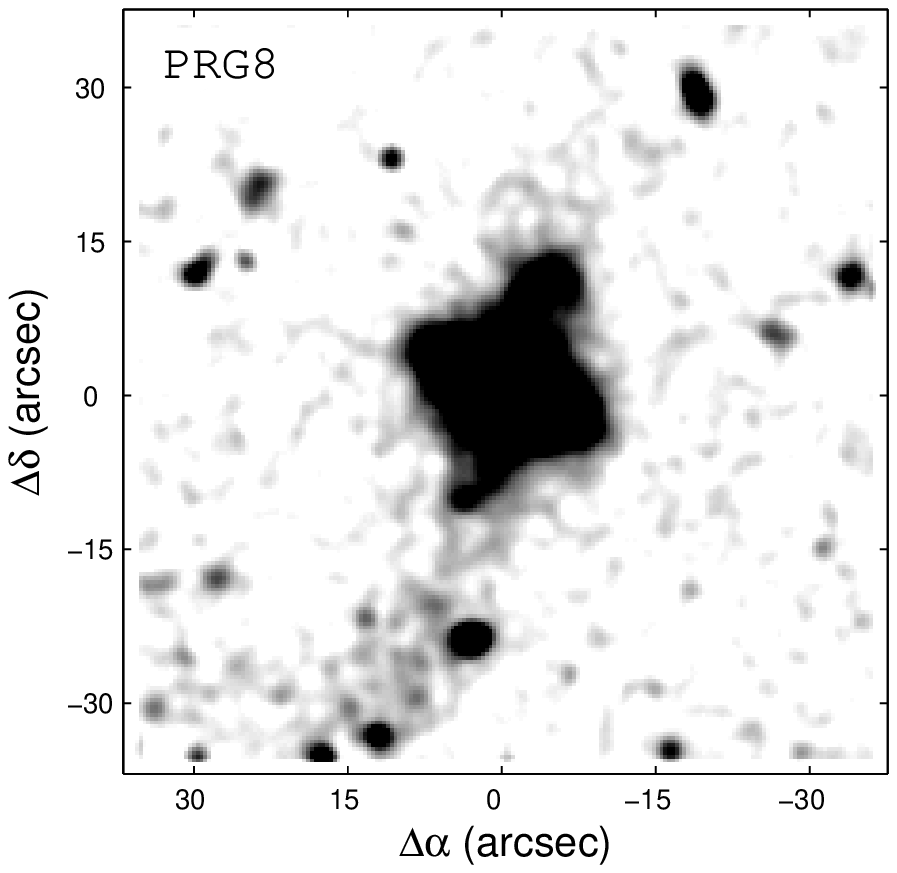} & \includegraphics[width=6cm]{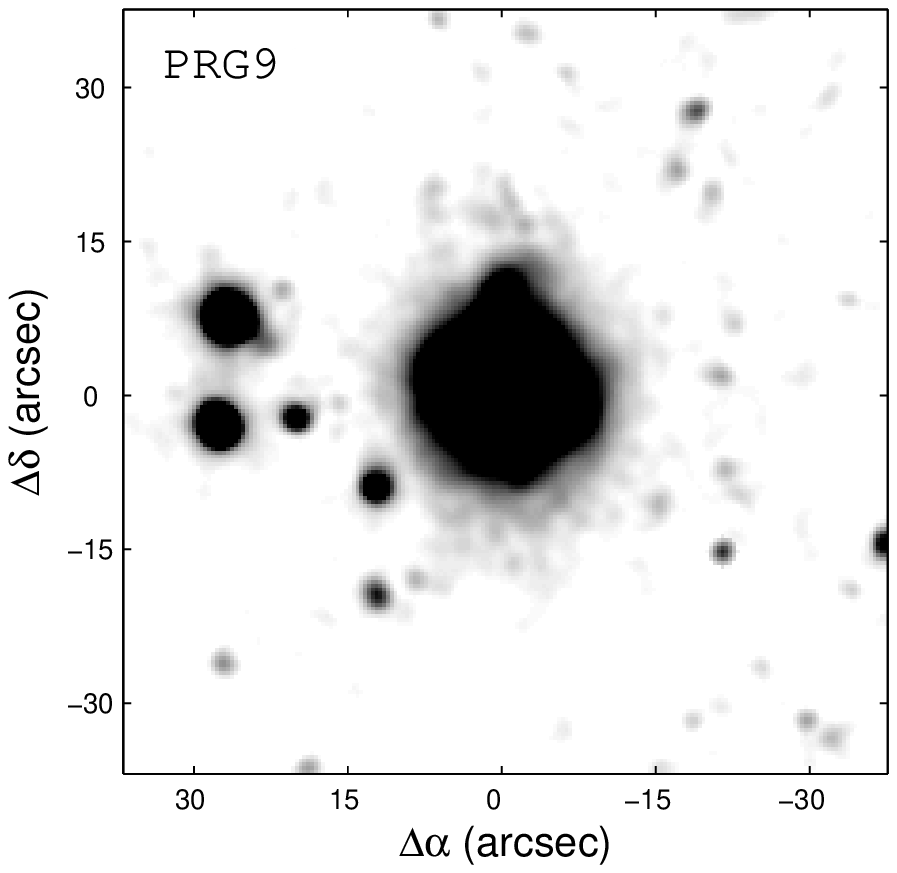}\\
\end{tabular}
\end{center}
\caption{$r$-band images smoothed over 3x3 pixels and displayed as negative to enhance low-level features. }
 \label{f:Rmaps}
\end{figure*}
\setcounter{figure}{2}
\begin{figure*}
\begin{center}
\begin{tabular}{ccc}
\includegraphics[width=6cm]{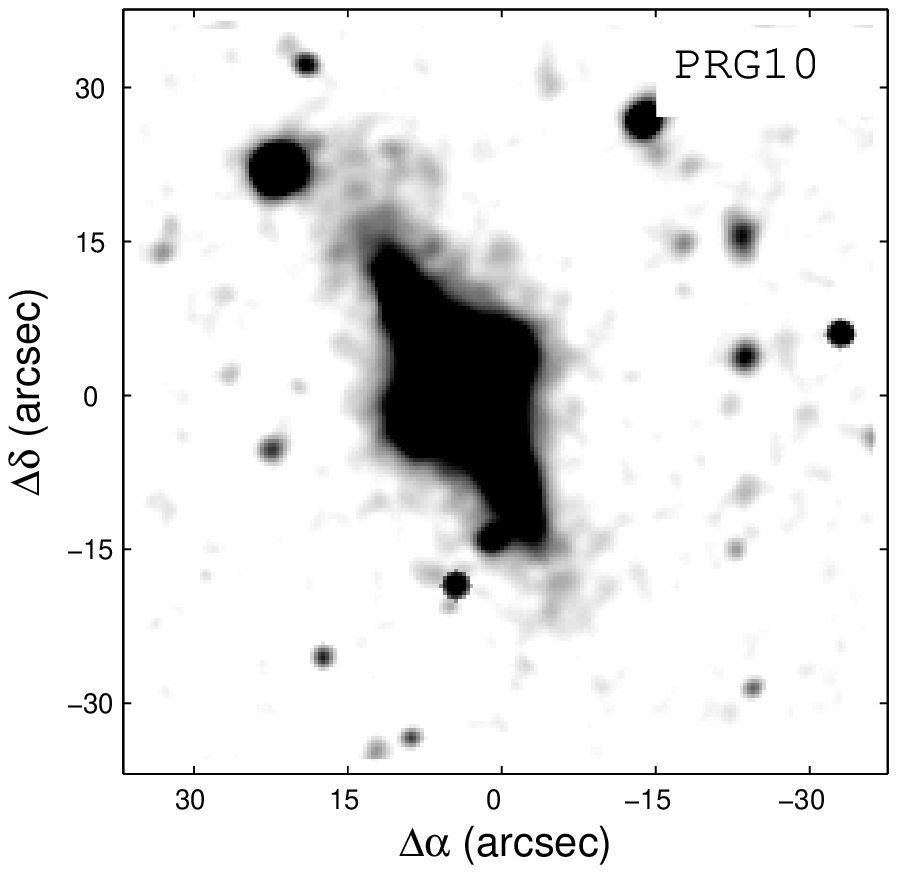} & \includegraphics[width=6cm]{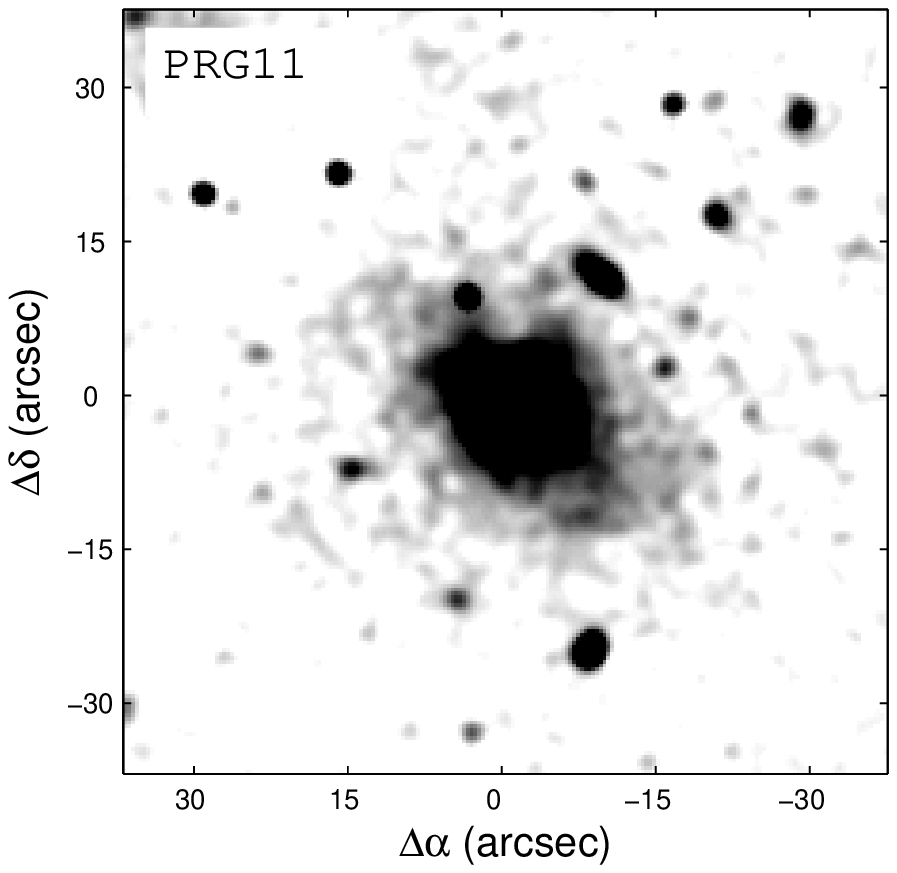} & \includegraphics[width=6cm]{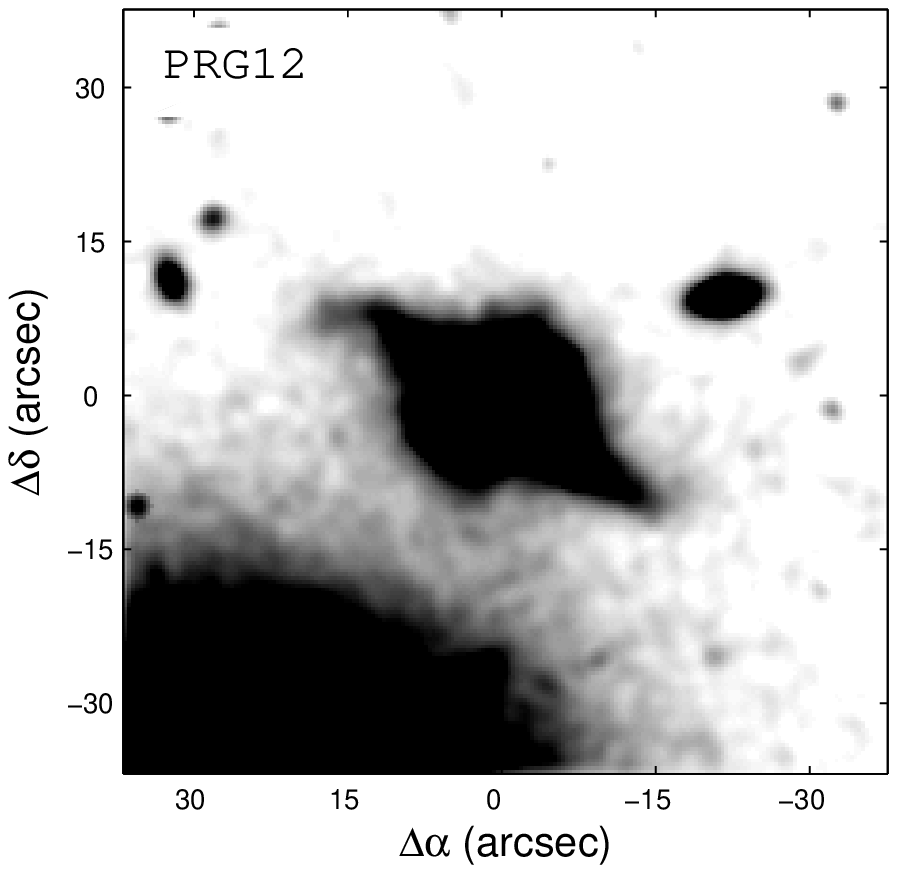} \\
 \vspace{-3mm}
\includegraphics[width=6cm]{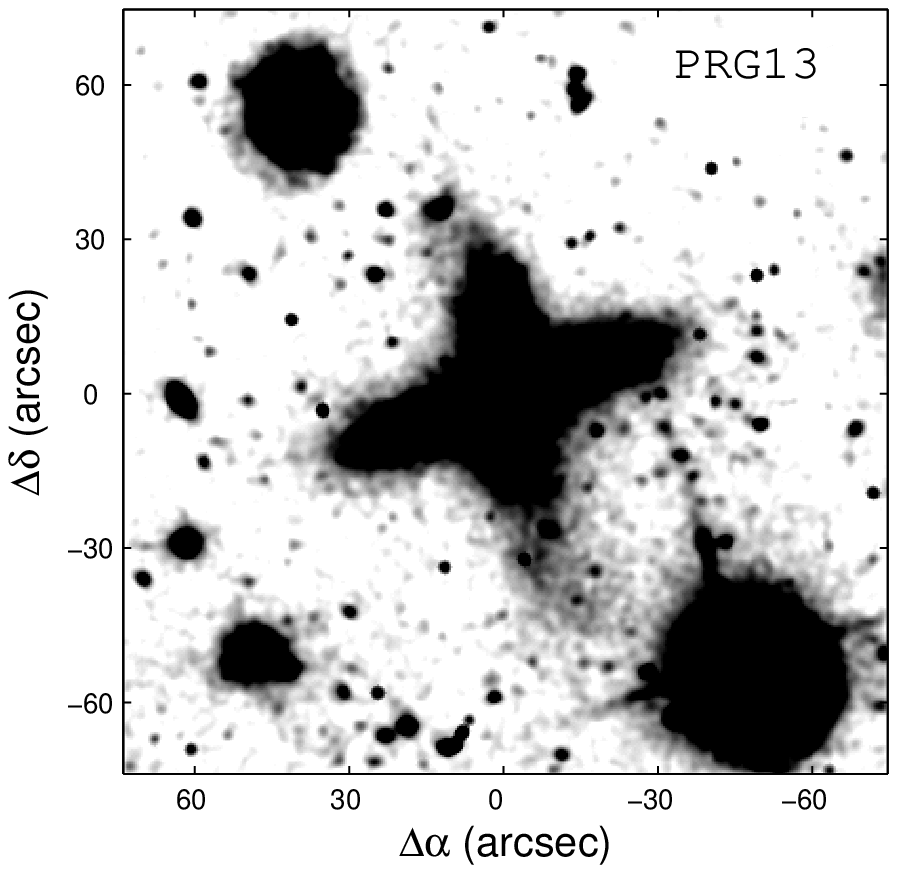} & \includegraphics[width=6cm]{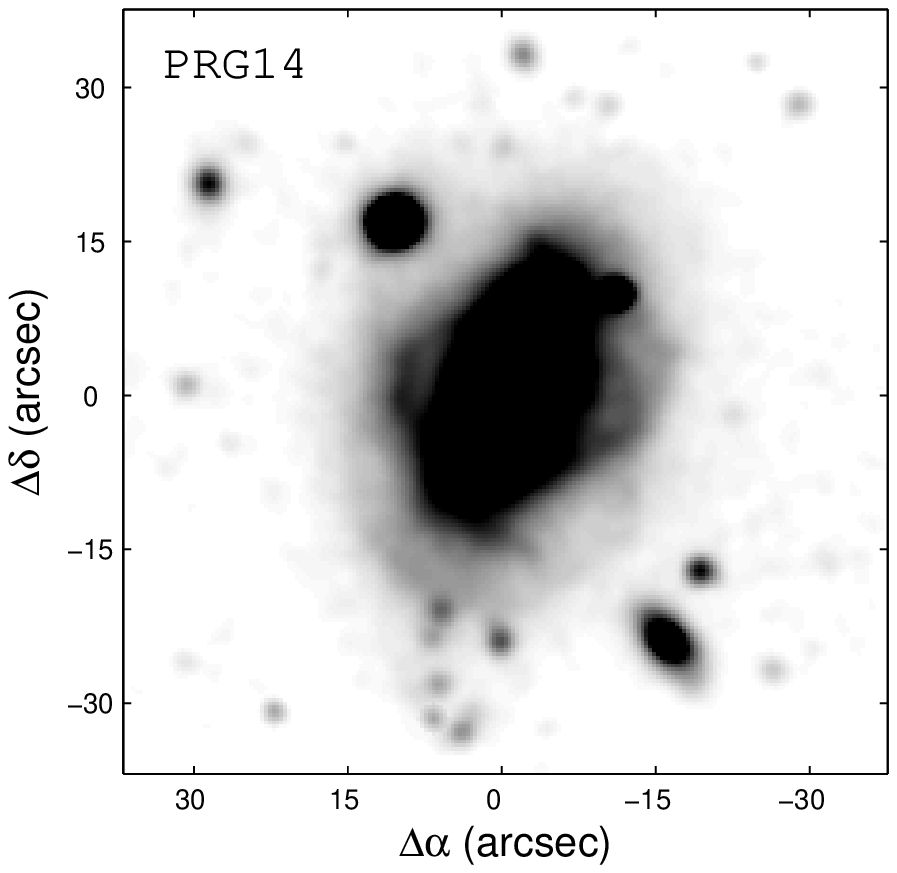} & \includegraphics[width=6cm]{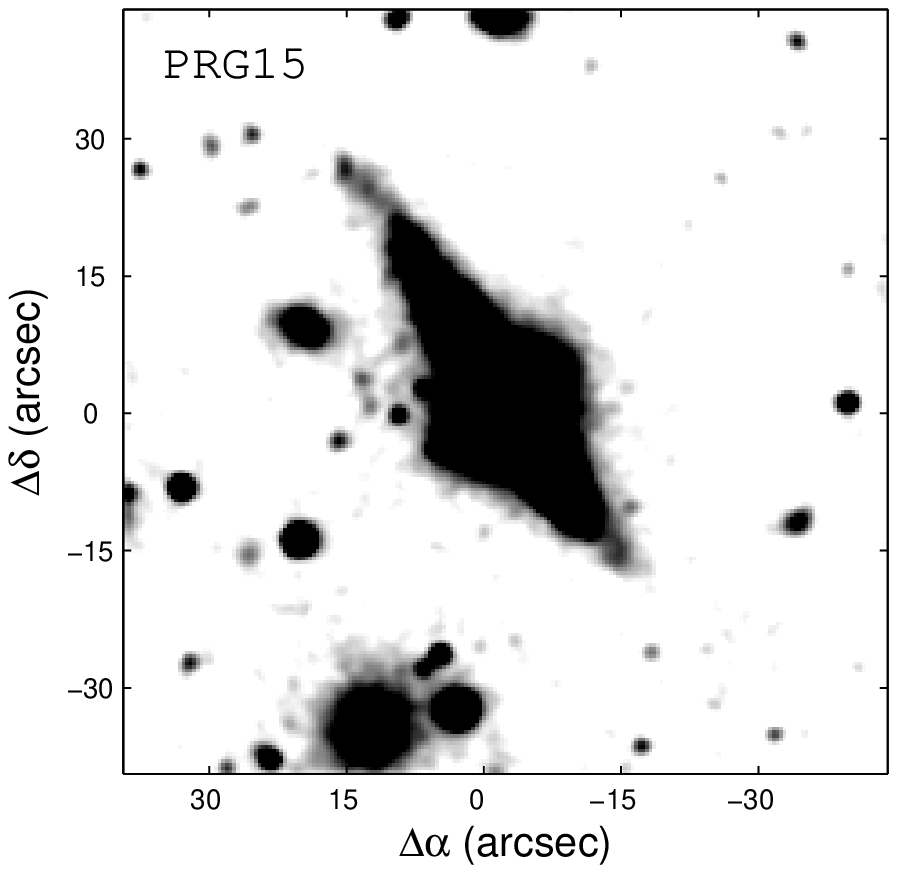}
 \\
 \vspace{-3mm}
\includegraphics[width=6cm]{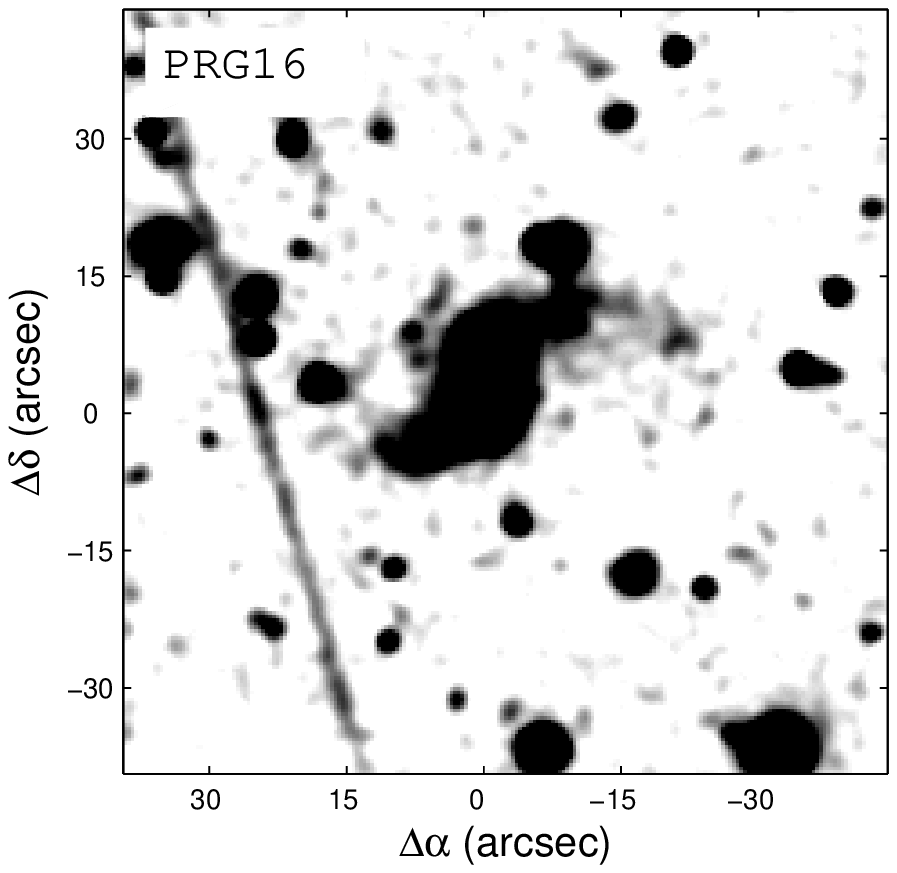} & {} & {} 
\end{tabular}
\end{center}
\caption{cont.}
 \label{f:}
\end{figure*}


\section{Individual galaxies}
\label{S:analysis}
In this section we discuss basic properties of our sample galaxies.
In particular, we focus on:
\begin{enumerate}[1.]
 \item The morphology of the main bodies and their surrounding structures. 
 \item The detection of extended low surface brightness features and fine structures in and around the galaxies. 
To enhance low-level features we display in Fig.\ 3 $r$-band images smoothed over 3x3 pixels.
 \item Examining the continuum-subtracted H$\alpha$+[NII] images to detect extended line-emission. 
 \item The analysis of SDSS spectra; the spectra obtained from the central 3 arcsec of each galaxy were measured by SDSS and are available through the MPA/JHU catalogue (see Brinchmann et al.\ 2004). 
The SDSS-III database and the MPA/JHU catalogues provide emission line classifications based on the Baldwin-Phillips-Terlevich (BPT) diagrams that distinguish active galactic nuclei (AGNs) from star-forming galaxies. 
 \item Studying the environment of each galaxy by searching the NASA/IPAC Extragalactic Database (NED) for neighbour galaxies within a projected distance of 2 Mpc and with recession velocities within 300 km sec$^{-1}$ of the value measured for the PRG.
\end{enumerate}
Note that a first follow-up spectral study of candidate PRGs by Moiseev et al.\ (2011) already confirmed the presence of a polar ring component in PRGs 4, 12 and 15.  \\

\noindent{\bf PRG 1 (SDSS J003209.80+010836.5):} \\
A puffy, diffuse disc is inclined by $\sim$20$^\circ$ to the polar axis of the main body. The deep Stripe82 image reveals a long tidal tail of faint luminous material which appears to connect to the polar ring at its northern side where a diffuse stellar source, the suspect donor, is detected. The tidal feature twists around the host and extends about one arcmin to the south. 
We find no other galaxies in the neighbourhood of PRG1. The SDSS spectrum shows significant central emission characteristic of a star-forming galaxy.

\noindent{\bf PRG 2 (SDSS J004812.18-001255.5):} \\
This peculiar galaxy lies close to the celestial equator and was imaged a number of times by the SDSS.
An asymmetric elongated structure crosses the main body at an angle of about 15$^\circ$ from the polar axis.
The arc-like structure is strongly bent and displaced relative to the nucleus, and shows a diffuse irregular clump at its east side.
The main body shows distorted outer isophotes and its surface brightness drops abruptly at the west side compared to the opposite side. 
A weak enhancement is observed at the north side of the host and faint diffuse light is detected outside the galaxy to its north-east.
The unsettled appearance of both the main body and the polar structure hints at a recent or ongoing gravitational interaction.
A close companion lies at a projected distance of about 1.6 arcmin ($\sim$100 kpc) from PRG 2.
However, the environment is not rich in galaxies and only one neighbour with a similar recession velocity is found within a 2 Mpc radius. 
The inner 3 arcsec region of the galaxy shows strong emission lines. 
The galaxy is classified as `composite' in the MPA/JHU catalogues, meaning that it contains significant contribution from an AGN and/or star formation activity.\\

\noindent{\bf PRG 3 (SDSS J024258.42-005709.3):} \\ 
Examining the publicly available SDSS colour image of this small galaxy reveals the presence of blue emission on both sides of the nucleus.
Light from these blue regions probably entered the SDSS fiber aperture and is likely related to the star formation activity detected by analyzing the prominent emission lines in the inner spectra.
The light from the central elliptical-like body dominates the inner isophotes whereas the faint polar structure is barely visible against the host  background starlight in single SDSS exposures.
We count at least 10 galaxies in the vicinity of PRG 3, with the nearest one at a projected distance of about 0.4 Mpc.
The images and central spectra of these neighbours indicate that star-forming galaxies are frequent in this environment.\\

\noindent{\bf PRG 4 (SDSS J082038.18+153659.7):} \\
The image of the galaxy shows the presence of a nearly edge-on disc with well-defined thin edges.
The angle between the polar disc and the minor axis of the host is about 5$^\circ$.
Closely inspecting the system reveals that the photocentre, defined by isophotes several arcsec from the nucleus, does not coincide with the the peak brightness location. This might be due to the apparent warping of the polar disc or due to an inclination effect.
The continuum-subtracted H$\alpha$+[NII] image reveals extended emission from the north-east part of the ring.
The SDSS spectrum shows absorption lines typical of late-type stars and weak (or no) emission lines. 
Five galaxies, including three starburst and star-forming galaxies, are found 1-2 Mpc away from PRG 4.\\

\noindent{\bf PRG 5 (SDSS J084832.00+322012.4):}\\
The appearance of the outer ring resembles to some extent that of PRG 4. The ring is tilted by about 20$^\circ$ with respect to the polar axis of the host galaxy, which is rounder than the main body of PRG 4. 
Diffuse light is detected close to the edge of the ring about 15 arcsec to the south-west. However, this red feature does not seem distorted and is possibly a background early-type galaxy.
PRG 5 is the most luminous member of a close group of at least four galaxies, whose spectra are characteristic of typical early-type galaxies. \\

\noindent{\bf PRG 6 (SDSS J091453.64+493824.0):}\\
This galaxy reveals a wealth of morphological detail characteristic of a disturbed merger remnant.
The outer isophotes of the main body are asymmetrically deformed and the surface brightness drops steeply towards north-east and south-west.  
A bright condensation is located $\sim$10 arcsec away from the centre of the galaxy at the west part of the ring and a sharp-edge arclet emerges at the east part of the ring about 18 arcsec from the nucleus.
Two thick elongated features extend more than 40 arcsec to the west and twice that distance to the east at a position angle that is intermediate between 
that of the host and the polar ring.
Furthermore, a small distorted object, is located about 20 arcsec north of the centre within a diffuse fan of faint material. 
In a merger view, this faint halo consists of stars that were dispersed by the collision but did not settle in the polar plane, while captured material has been oscillating at large and small distances from the central body (Bournaud \& Combes 2003). 
We found four galaxies of similar luminosities in the neighbourhood of PRG 6, two of which are classified as star-forming by SDSS-III.
No significant emission is detected in either our continuum-subtracted H$\alpha$+[NII] image or the SDSS spectra.\\

\noindent{\bf PRG 7 (SDSS J094207.34+362417.2):}\\
The elliptical body is crossed by an asymmetrical stellar disc tilted $\sim$15$^\circ$ from the polar axis. 
Although the north-east part of the ring seems to end with a faint light enhancement, a careful inspection reveals diffuse light stretching farther away at this direction. The opposite side of the ring is relatively relaxed.
Two galaxies, a bluish star-forming galaxy and an early-type galaxy, are found at projected distances of about 1 Mpc from PRG 7. \\

\noindent{\bf PRG 8 (SDSS J094302.33-004850.0):} \\
At a first glance, the S0 host appears on the single SDSS frames to be surrounded by a warped disc. Our deep images show  
that this structure is probably a fragmented and highly inclined ring. The ring extends more than 
twice the radius of the host,  but has a faint thick tail stretching about 40 arcsec to the south-east.
The isophotes of the main body become more eccentric at larger radii, implying a more substantial disc than for other hosts in our sample.
NED lists four reddish galaxies with known redshift near PRG 8, the closest one at a projected distance of about 0.3 Mpc.
The SDSS spectrum of PRG 8 shows the characteristic signature of a typical Sy2 nucleus.  \\
   
\noindent{\bf PRG 9 (SDSS J104623.65+063710.2):} \\
The isophotes of the galaxy are disturbed by internal dust extinction. 
A faint asymmetric structure crosses the galaxy $\sim$25$^\circ$ from the minor axis and is almost hidden by the outer isophotes of the main body. 
Our continuum-subtracted H$\alpha$+[NII] image reveals significant emission 
in the central parts of the galaxy and along the polar structure. 
The SDSS fiber also detected strong emission lines characteristic of starburst galaxies. 
The environment is rich in galaxies, several of them 2-3 mag brighter than PRG 8.
The vast majority of these neighbours are star-forming or starburst galaxies. \\
 
\noindent{\bf PRG 10 (SDSS J114444.02+230944.9):} \\
This is the best candidate of a polar ring galaxy in our sample since the two perpendicular components are bright and well-defined.
The ring extends far beyond the host and is slightly warped at the edges. Our deep images reveal a weak dust lane following the ring at the north-west 
side of the host and a faint diffuse halo surrounding the north-east edge of the ring. 
The continuum-subtracted H$\alpha$+[NII] image shows extended emission along the ring, indicating the presence of recently formed stellar population. 
The light in the inner 3 arcsec of the galaxy does not show clear signs of star formation and is characteristic of early-type galaxies.
The galaxy belongs to a compact group of several galaxies distributed over 0.4 Mpc projected distance. \\

\noindent{\bf PRG 11 (SDSS J115228.29+050044.7):} \\
The faint thick disc, which crosses the main body at $\sim$5$^\circ$ from polar axis, is very similar to that of PRG 3. 
Our deep images show broad fans of diffuse light around the galaxy, and an elongated feature 15 arcsec from nucleus on the north-west side of the ring. However, we cannot determine whether this is stellar debris gravitationally bound to the system or light from a background source.
The SDSS spectrum shows strong emission lines but the analysis of their nature is not decisive, and the object is classified `composite' in the MPA/JHU catalogue. The combination of strong narrow components and a weak broad H$\alpha$ component implies that both star formation and nuclear activity contribute significantly to the spectrum (Wang \& Wei 2008).
The galaxy lies in a dense environment and is located about 1.3 Mpc in projected distance from CGCG 040-053, an elliptical galaxy listed in NED as the brightest galaxy of its cluster. The three closest neighbours to PRG 10 are identified in SDSS-III as star-forming galaxies. \\

\noindent{\bf PRG 12 (SDSS J130816.92+452235.1):}\\ 
The disc is inclined $\sim$30$^\circ$ to the minor axis of the host, extends to about three times the radius of the host and exhibits internal dust obscuration which slightly distorts the outer isophotes of the host. 
The north-east edge of the ring is twisted by $\sim$45$^\circ$ to the east.
Our continuum-subtracted H$\alpha$ image shows some emission from the host, but this is barely, if at all, seen in the SDSS spectrum. 
We count at least 15 galaxies within 2 Mpc around PRG 12. 
The brightest galaxy in this neighbourhood lies only one arcmin (about 40 kpc in projected distance) from PRG 12, but the two galaxies do not show clear signs of interaction. \\

\noindent{\bf PRG 13 (SDSS J135941.70+250046.0):} \\
The main body of this galaxy is unusual. The high eccentricity, prominent dust lane along the major axis and the apparent warped structure indicate that this is a highly-inclined disc warped on both sides.
Our deep images reveal an asymmetric polar ring 41 arcsec wide, tilted $\sim$15$^\circ$ from the minor axis.
The ring extends far beyond the main body and contains several luminous blue clumps. 
A warped diffuse outer structure extends beyond the ring and is too faint to be seen in single SDSS exposures.
Our continuum subtracted H$\alpha$+[NII] image shows no emission along the ring. The central parts of the host are badly subtracted in our images but the central 3 arcsec region sampled by the SDSS fiber shows no prominent emission lines.
The neighbourhood of PRG 13 includes both a compact group of galaxies (no.\ 205 in the UZC catalog, Focardi \& Kelm 2002) and a galaxy classified as isolated (CGCG 132-069, see Karachentseva 1973). \\

\noindent{\bf PRG 14 (SDSS J151114.08+370237.6):} \\
The SDSS colour images (see Moiseev et al.\ 2011) show an inclined bluish ring around the central body. However, our deep images reveal a more complex structure where the surrounding material appears to form more than one complete turn around the galaxy, as in NGC 2685. Note that what we interpret as an helical structure around the host might actually be diffuse stellar fans on both sides of the main body. 
PRG 14 host an AGN and is a member of a compact group of several early-type galaxies. \\

\noindent{\bf PRG 15 (SDSS J204805.66+000407.8):} \\
The recession velocity of this object is the lowest in our sample.
The peculiar appearance of this object was noted already by Zwicky \& Zwicky (1971) and the extended, luminous polar structure can be seen clearly even on single SDSS frames.
The dim light within an outer bright rim gives the polar structure a ring-like appearance. The ring light does not drop abruptly to zero on both sides of the ring and a faint tail extends from the north-east side of the ring.
Strong internal dust extinction is detected along the north-west side of the ring.
Ionized gas is detected in the continuum-subtracted H$\alpha$+[NII] image of the galaxy as several emission clumps along the ring. 
A close star-forming companion lies about three arcmin (about 900 kpc) north of PRG 15. Several other galaxies, mostly blue and star-forming, are found at different distances around the galaxy. \\

\noindent{\bf PRG 16 (SDSS J212339.15-002235.2):} \\
This galaxy is an example of an object with faint extended structure that could not have been detected in single SDSS exposures. 
Stacking the multiple SDSS images, taken as part as the Stripe82 project, reveals two arm-like features spiralling out from both sides of the main body. 
PRG 16 does not resemble classical PRGs and is the least promising PRG candidate in our sample. In fact, it is the only object in our sample classified in the SPRC as a `related object', rather than a `good' or `best' candidate.
The SDSS spectrum of PRG 16 shows significant emission characteristic of star formation activity.
The galaxy has a nearby companion at a projected distance of about 0.4 Mpc, and several other neighbours 1.5-2.0 Mpc away.
Most of these neighbours are rather blue and show significant central emission indicative of AGN or star formation activity. \\

\section{Discussion}
\label{S:discuss}
To study the basic morphological properties and optical colour of the hosts and polar rings we observed 11 candidate PRG in $u$ and $r$ and used archive Stripe82 data for five additional probable PRGs. 
We note that some polar rings galaxies are very faint and are hardly visible on standard SDSS images.
In fact, the polar structures around PRGs 2, 3 and 16 were detected on the deep Stripe82 images rather than in standard SDSS frames, which highlights the importance of deep galaxy surveys in identifying such objects and determining their frequency (see also Finkelman, Graur \& Brosch 2011).
We discuss below some of the basic properties of PRGs and compare them with other types of early-type galaxies (ETGs).

\subsection{Morphology}
\begin{figure*}
\begin{center}
\begin{tabular}{cc}
 \includegraphics[width=8cm]{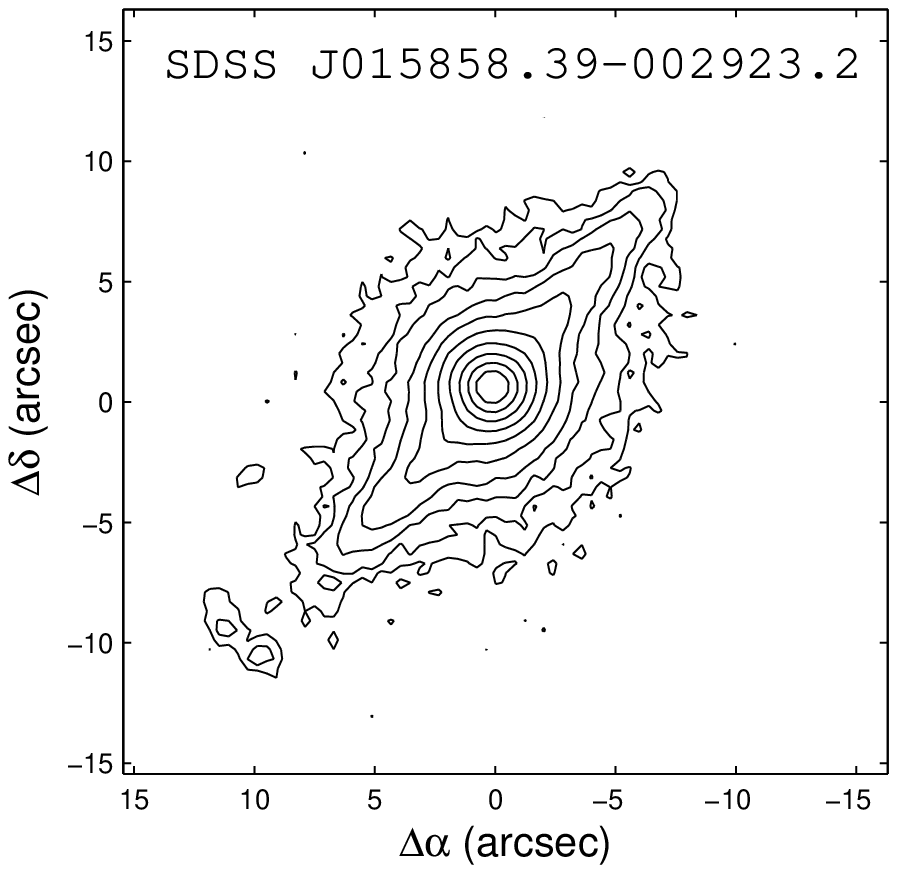} &  \includegraphics[width=8cm]{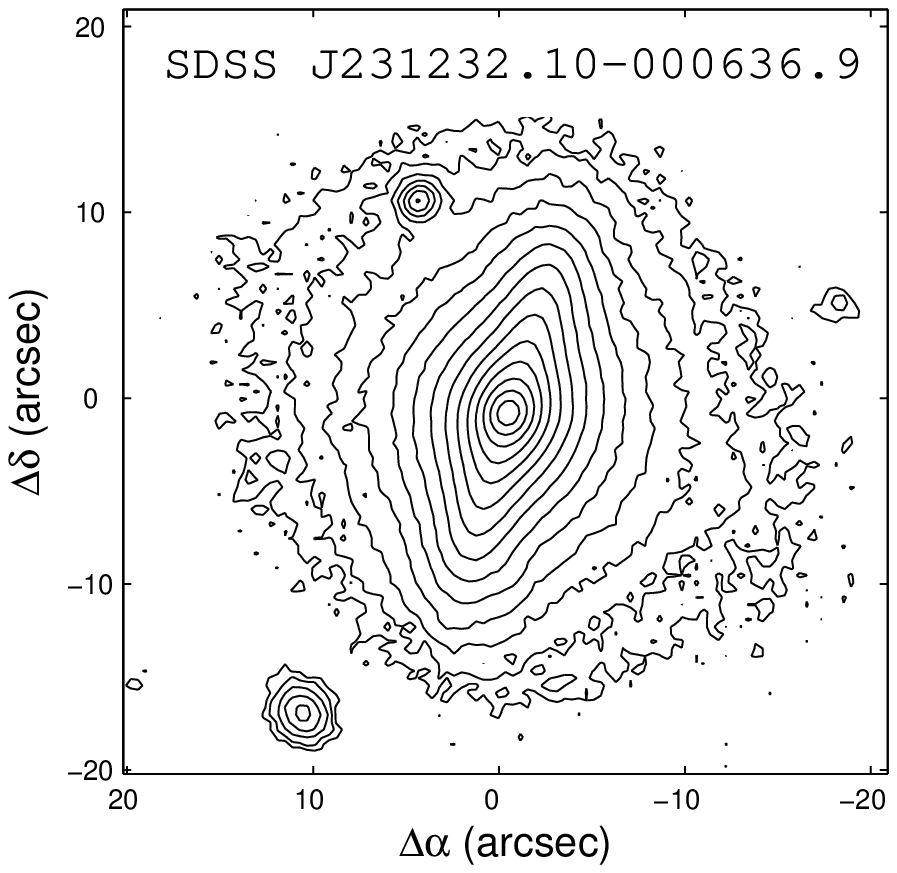}\\
\end{tabular}
\end{center}
\caption{SDSS J015858.39-002923.2 and SDSS J231232.10-000636.9, objects number 77 and 185 in the SPRC, resemble `normal' disc galaxies with a symmetric bright disc around their central bulge.}
 \label{f:ex}
\end{figure*}

Our sample galaxies show a wide variety of morphological properties including systems with extended disc-like rings
with central `holes' and galaxies with a dominant bulge and a relatively narrow ring which is not extended (see also Reshetnikov \& Stanikova 1997). In fact, some of the galaxies look similar to classical PRGs. 
For instance, PRG 10 shows an extended structure similar to NGC 4650A and PRG 15 resembles A0136-0801.
SDSS J015858.39-002923.2 and SDSS J231232.10-000636.9, objects number 77 and 185 in the SPRC, are included also in the Stripe82 database but are not part of our sample. Fig.\ \ref{f:ex} shows that both these objects resemble `normal' disc galaxies with a symmetric bright disc around their central bulge rather than typical PRGs. However, note that the deep coadded image of SPRC 185 shows an extended stellar envelope extending in a direction perpendicular to the disc plane.

By careful visual inspection of the optical images we detect diffuse low surface brightness material distributed around some of our systems. 
Asymmetric tails along the ring axes (PRGs 6, 7 and 8), disrupted stellar objects (PRGs 2 and 5) and broad stellar fans (PRGs 11 and 14) are likely signatures of recent interactions. Several galaxies exhibit sharp-edge features which presumably formed by infalling material oscillating at alternating radii before gathering in the ring (PRG 6 is the best example).

Both merger and gas accretion events can account for the formation of polar rings around pre-existing galaxies (Bekki 1997; Reshetnikov \& Sotnikova 1997; Bournaud \& Combes 2003). 
However, numerical simulations predict significant morphological and dynamical differences between the results of both scenarios. 
For instance, the material transfered from the donor galaxy to the host during accretion consists mostly of gas which settles in the polar plane while fueling star formation and subsequently forming a luminous polar structure, whereas during galaxy collisions part of the stars disperse and form a diffuse stellar halo around the PRG.
The formation of the ring in each scenario depends also on physical parameters, such as the masses of the interacting galaxies. In the merging scenario a certain range of mass ratio of galaxies is required to form a polar ring galaxy (Bekki 1997, 1998), whereas external accretion of gas can form not only rings significantly lighter than the host galaxies (Iodice et al.\ 2002), but also rings with masses comparable to the host masses (Bournaud \& Combes 2003).
Bournaud \& Combes (2003) further argued that the accretion scenario is much more probable than the merging scenario, thus many faint polar rings have probably not yet been detected. The detection of a significant number of new candidate PRGs in the Galaxy Zoo seems to provide circumstantial evidence to this view.  

\subsection{Optical colour}
\begin{figure*}
\begin{center}
\begin{tabular}{cc}
 \includegraphics[width=8cm]{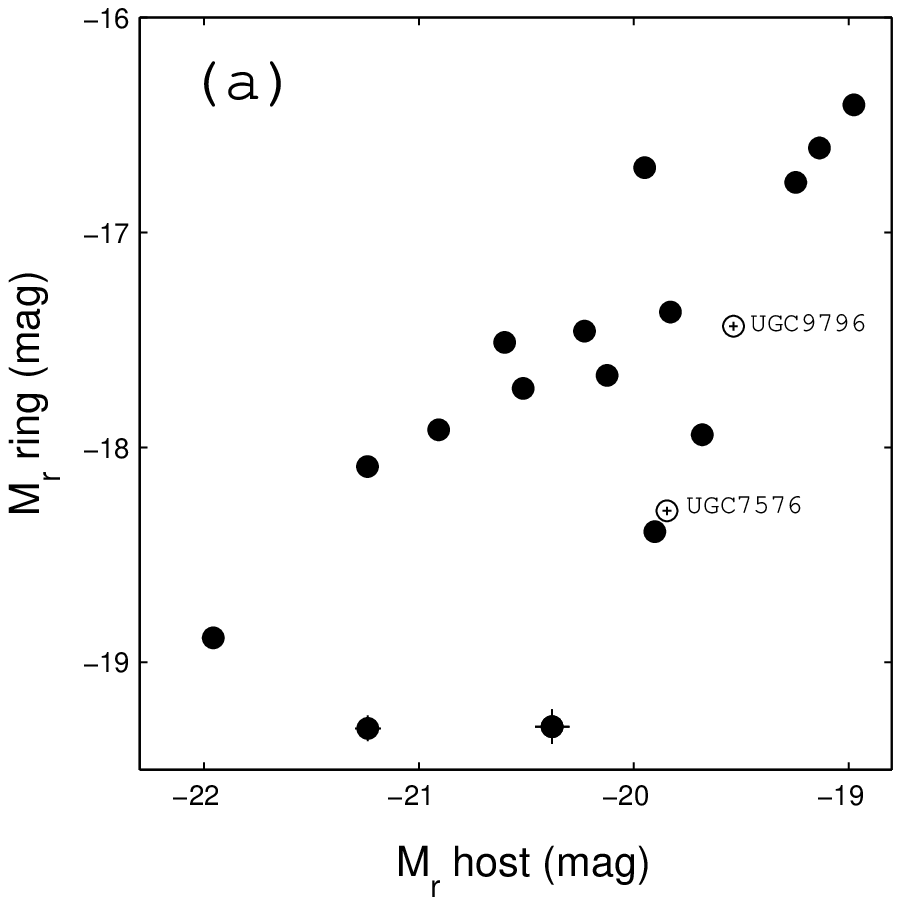} &  \includegraphics[width=8cm]{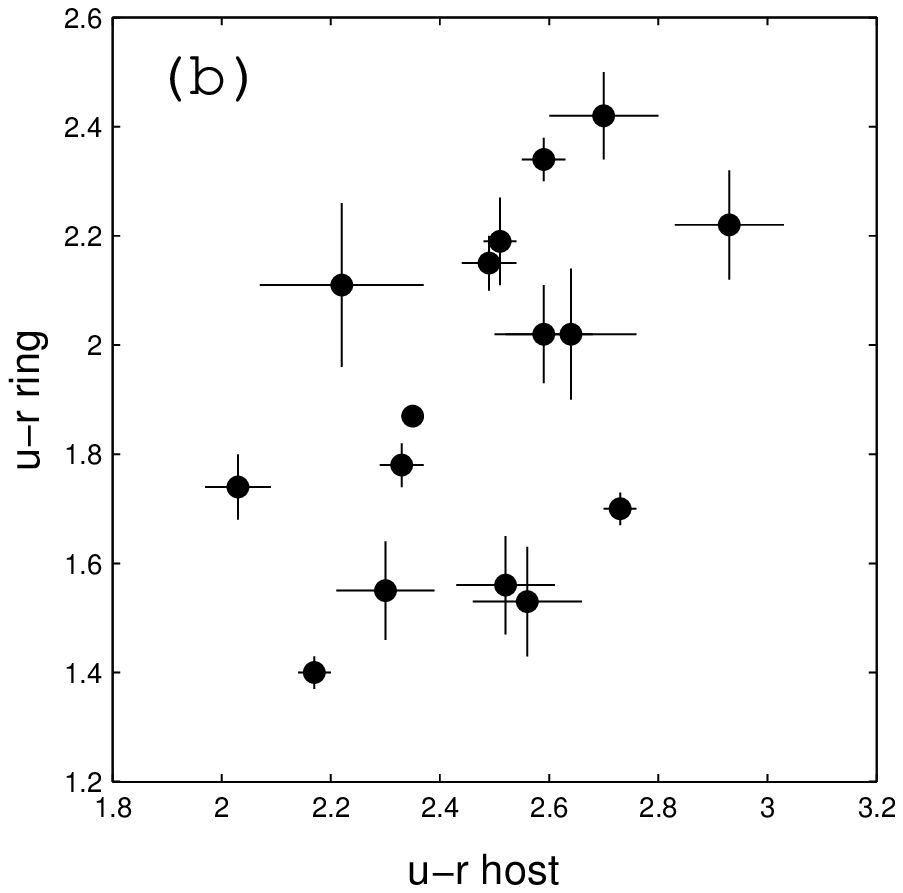}\\
 \includegraphics[width=8cm]{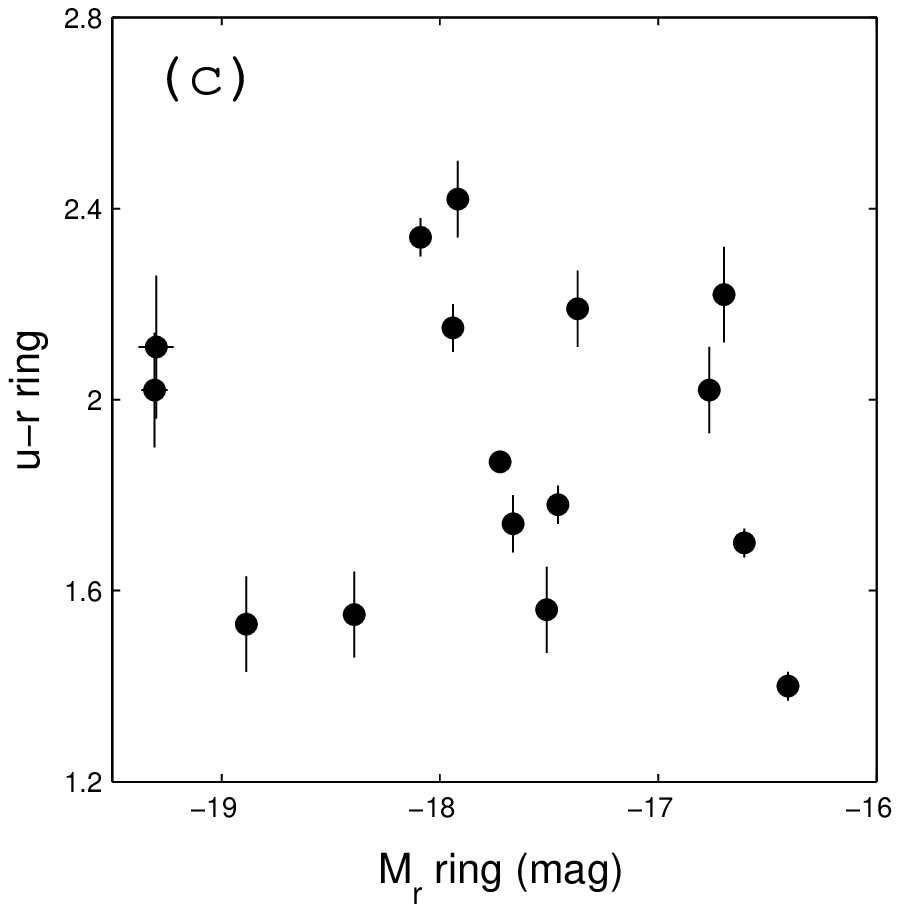} &  {}\\
\end{tabular}
\end{center}
\caption{Upper left panel: absolute $r$-magnitude values of hosts versus rings. The plot includes also the values measured from SDSS images for two known PRGs, UGC 7576 and UGC 9796. Upper right panel: $u$-$r$ colours of hosts versus rings. Lower left panel: absolute $r$-magnitude versus $u$-$r$ colour of rings. }
 \label{f:panels}
\end{figure*}

To study the relation between the two decoupled stellar components we plot in panel a of Fig.\ \ref{f:panels} the absolute $r$-magnitudes of the host galaxies versus the polar rings (see also Table \ref{t:colour}).
We find that the light of the ring is about 2-3 magnitudes fainter than that of the host, and that the two measures seem to correlate over almost three orders of magnitude.
Although we know that several genuine PRGs host rings as massive as the hosts, whether our result is related to fundamental processes or to a bias of our sample selection is yet unclear. Explaining our result as an outcome of a certain formation channel could be problematic; diffuse stellar halos, a distinctive signature of mergers, are visible in our deep images only around some of the systems.

We find also that all hosts have overall colours redder than the polar structures. 
These characteristics seem to describe also the majority of PRGs in previous studies (Iodice et al.\ 2002; Godinez-Martinez et al.\ 2007).
As shown in panel b of Fig.\ \ref{f:panels}, the $u$-$r$ colour of the hosts and the rings seem to correlate, providing yet another clue to the nature of the relation between the stellar content of these two components.   
We caution, however, that this observed relation can partly be due to the bluish light emitted from the highly-inclined luminous ring which is projected against the host body and is not properly subtracted. As shown in panel c of Fig.\ \ref{f:panels}, no relation is found between the luminosity of the polar structure and its colour. 


\begin{table*}
 \centering
  \caption{Photometric properties of galaxies in our sample.
  \label{t:colour}}
\begin{tabular}{rrccrcc}
\hline
{} &  \multicolumn{3}{l}{{\bf Host:}} & \multicolumn{3}{l}{{\bf Ring:}}\\
PRG \#  &   PA ($^\circ$) & $r$ (mag) & ($u$-$r$) & PA ($^\circ$) & $r$(mag) & ($u$-$r$)         \\
\hline
{1} & 175 & $16.17\pm0.06$ & $2.22\pm0.15$ & 105 &$17.75\pm0.06$ & $2.11\pm0.15$ \\
{2} & 162 & $16.43\pm0.01$ & $2.35\pm0.02$ & 88 &$19.22\pm0.01$ & $1.87\pm0.02$ \\
{3} & 20 & $17.35\pm0.01$ & $2.17\pm0.03$ & 120 & $19.92\pm0.01$ & $1.40\pm0.03$  \\
{4} & 138 & $16.35\pm0.03$ & $2.93\pm0.10$ & 43 & $19.60\pm0.03$ & $2.22\pm0.10$ \\
{5} & 123 & $16.06\pm0.06$ & $2.64\pm0.12$ & 34 & $17.99\pm0.06$ & $2.02\pm0.12$ \\
{6} & 15 & $15.52\pm0.02$ & $2.03\pm0.06$ & 100 & $17.98\pm0.02$ & $1.74\pm0.06$ \\
{7} & 126 & $17.26\pm0.04$ & $2.51\pm0.03$ & 52 & $19.72\pm0.04$ & $2.19\pm0.08$ \\
{8} & 63 & $16.12\pm0.02$ & $2.59\pm0.04$ & 157 & $19.27\pm0.02$ & $2.34\pm0.04$ \\
{9} & 68 & $16.21\pm0.02$ & $2.73\pm0.03$ & 5 & $18.74\pm0.02$ & $1.70\pm0.03$  \\
{10} & 125 & $16.62\pm0.03$ & $2.30\pm0.09$ & 31 & $18.13\pm0.03$ & $1.55\pm0.09$  \\
{11}& 53 & $17.07\pm0.02$ & $2.52\pm0.09$ & 138 & $20.16\pm0.02$ & $1.56\pm0.09$  \\
{12}& 117 & $16.22\pm0.03$ & $2.59\pm0.09$ & 55 &$18.70\pm0.03$ & $2.02\pm0.09$  \\
{13}& 110 & $14.70\pm0.03$ & $2.70\pm0.10$ & 4 & $17.69\pm0.03$ & $2.42\pm0.08$ \\
{14}& 157 & $14.93\pm0.05$ & $2.56\pm0.10$ & 79 & $18.00\pm0.05$ & $1.53\pm0.10$  \\
{15}& 129 & $15.40\pm0.01$ & $2.49\pm0.05$ & 32 & $17.14\pm0.01$ & $2.15\pm0.05$ \\
{16}& 5 & $16.93\pm0.03$ & $2.33\pm0.04$ & 137 & $19.70\pm0.03$ & $1.78\pm0.04$  \\
\hline
\end{tabular}
\end{table*}

A visual inspection of the inner isophotes suggests that the main body in all our PRGs excluding PRG 13, is similar to an early-type galaxy, most often an S0.
We examined publicly available SDSS spectra of the inner 3 arcsec region of each galaxy and found in all cases the SDSS continuum and the absorption line spectra to be typical of a population of late-type stars.

\begin{figure*}
\begin{center}
\begin{tabular}{cc}
 \includegraphics[width=8cm, trim = 0cm 0 8cm 0, clip]{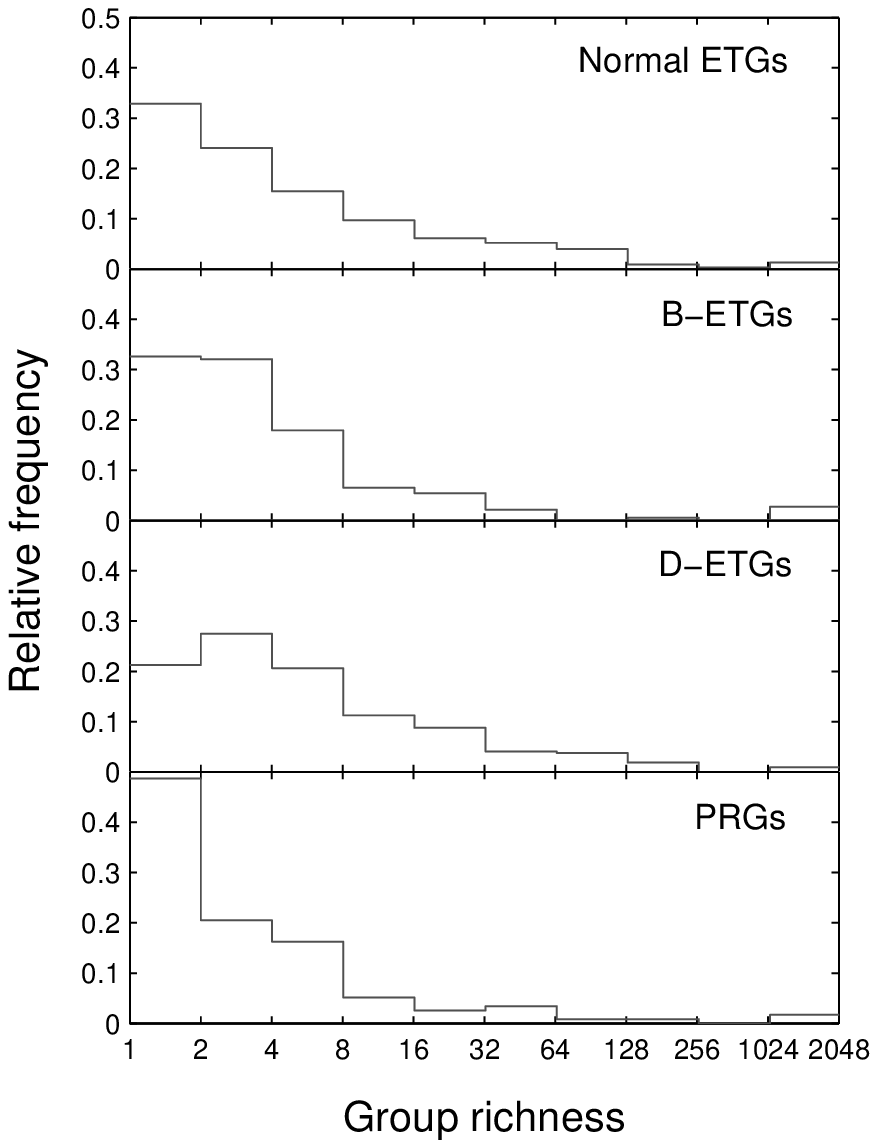} &  \includegraphics[width=8cm, trim = 0cm 0 8cm 0, clip]{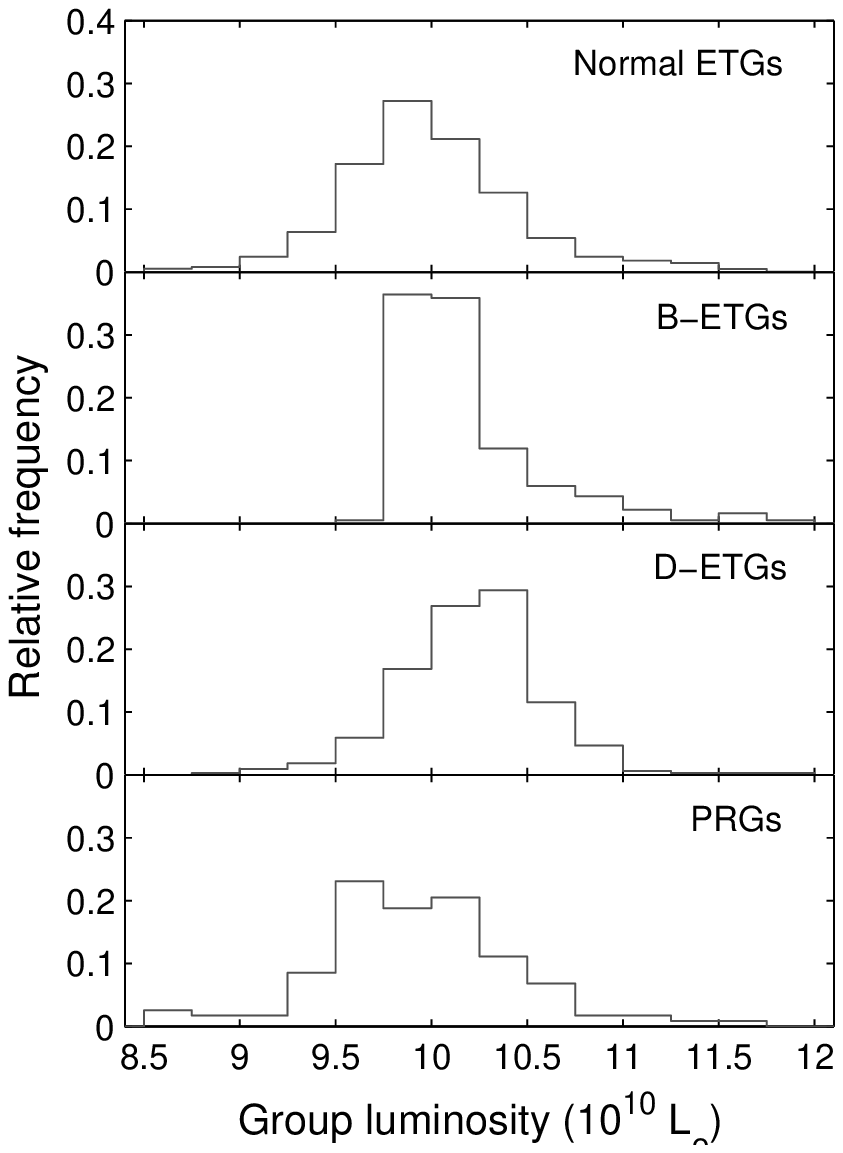}\\
 \includegraphics[width=8cm, trim = 0cm 0 8cm 0, clip]{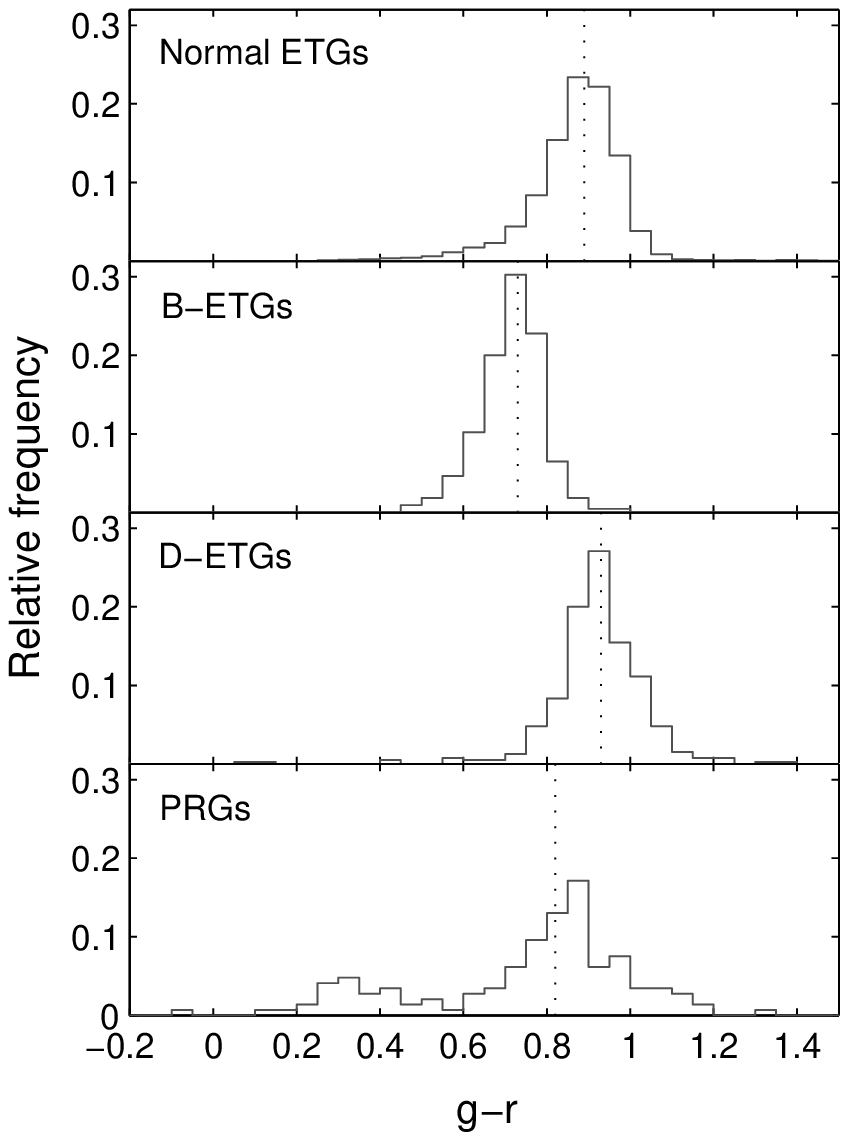} &  {}\\
\end{tabular}
\end{center}
\caption{Optical $g$-$r$ colour, group luminosity and group richness distributions of different classes of galaxies. 
Median colour values are shown using the dashed vertical lines.
`Normal' ETGs are selected from Galaxy Zoo 1, B-ETGS from Schawinski et al.\ (2009), D-ETGs from Kaviraj et al.\ (2011) and PRGs from Moiseev et al.\ (2011).}
 \label{f:compare}
\end{figure*}

In order to draw more general conclusions about the evolution of PRGs we compare the optical colour of candidate PRGs from the SPRC with that of a control sample of galaxies from the general early-type galaxies population in Galaxy Zoo 1 (hereafter `normal' ETGs);
these galaxies were selected under the criteria that they were classified by at least 50 Galaxy Zoo volunteers and that more than 70\% of the users flagged each object as an early-type galaxy. 
We then selectively deleted galaxies so that the control sample, consisting of $\sim$15 000 galaxies, would have the same redshift and magnitude distributions as the PRG population (see Kaviraj et al.\ 2011).
For comparison, we obtain also the colours of optically blue early-type galaxies (hereafter B-ETGs; Schawinski et al.\ 2009) and of dusty early-type galaxies (hereafter D-ETGs; Kaviraj et al.\ 2011, private communication).
As shown in Fig.\ \ref{f:compare}, the colour distribution of PRGs exhibits a clear bimodality with most objects showing $g$-$r$ colours similar to those of `normal' ETGs whereas several other PRGs are even bluer than the B-ETGs. This result can be explained if the `blue' PRG population represents late-type hosts, which are relatively rare among such systems. Internal dust obscuration naturally explains why D-ETGs tend to be redder than `normal' ETGs.

\subsection{Emission line classification}
The SDSS spectroscopic measurements of most sample galaxies show strong line emission from ionized gas.
The MPA/JHU catalogues provide emission line classification based on the BPT diagram using the methodology described in Brinchmann et al.\ (2004). Galaxies are divided into `star-forming' (SF), `composite', `AGN', `low S/N SF', `low S/N low-ionization nuclear emission-line region (LINERs)', and `unclassifiable' (galaxies with no or very weak emission lines) categories.
Among our objects three were confidently detected as AGN hosts, four other show clear evidence of recent or ongoing star formation activity and two galaxies exhibit a composite starburst/AGN activity. 
The BPT classification of galaxies, including those with low S/N emission-lines, is listed in Table 1. 

If polar rings are formed during galaxy interactions we would expect PRGs to have a better chance of hosting non-thermal nuclear activity or starbursts than the general population of early-type galaxies.
A comparison of the central activity in PRGs, B-ETGs, D-ETGs and `normal' ETGs is presented in Table \ref{t:compare}.
We find that 26\%-36\% of the `best' and `good' candidates in the SPRC host an active nucleus while 58\%-74\% of the galaxies are star-forming. The first numbers include confident classifications and `composite' galaxies, and the second ones include also galaxies with low S/N emission lines.
This analysis shows that the fraction of active nuclei among PRGs and PRG candidate is not significantly higher compared to `normal' ETGs. This is inconsistent with the result by Reshetnikov, Fa\'{u}ndez-Aban \&, de Oliveira-Abans (2001) who found that 41\%-63\% of their sample of 27 galaxies host LINER or Seyfert nuclei.

\begin{table*}
 \centering
  \caption{Comparison of star formation and nuclear activity, derived
using emission line classification based on the BPT diagram optical emission-line-ratio diagnostics using the methodology described in Brinchmann et al.\ (2004). `Normal' ETGs are selected from Galaxy Zoo 1, B-ETGS from Schawinski et al.\ (2009), D-ETGs from Kaviraj et al.\ (2011) and PRGs from Moiseev et al.\ (2011).
  \label{t:compare}}
\begin{tabular}{lrrrr}
\hline
     & PRGs  & D-ETGs & B-ETGs & Control ETGs \\ 
\hline
SF             & 38\% & 9\%  & 28\%      & 6\% \\
Composite      & 20\% & 16\% & 23\%      & 6\% \\
AGN            & 6\%  & 27\% & 16\%      & 4\% \\
Low S/N LINER  & 10\% & 34\% & 6\%       & 13\% \\
Low S/N SF     & 16\% & 8\%  & 7\%       & 15\% \\
Unclassified   & 10\% & 6\% & 20\%      & 56\% \\
\hline
\end{tabular}
\end{table*}
Starbursts are more frequent among PRGs relative to any other type of galaxy in our comparison.
Note that the SDSS fiber covers only the inner 3 arcsec of a galaxy, whereas emission-line imaging allows mapping the distribution of ionized gas throughout it. We detected clear evidence of extended H$\alpha$+[NII] emission in at least 4 out of 7 of our sample galaxies (PRG 4, 9, 10 and 15), implying that they have experienced a recent star formation episode. Not all of these objects show clear signs of starburst in their central spectra, which indicates that the fraction of star-forming PRGs deduced from the BPT classification is only a lower limit. 

\begin{figure}
\begin{center}
 \includegraphics[width=8cm]{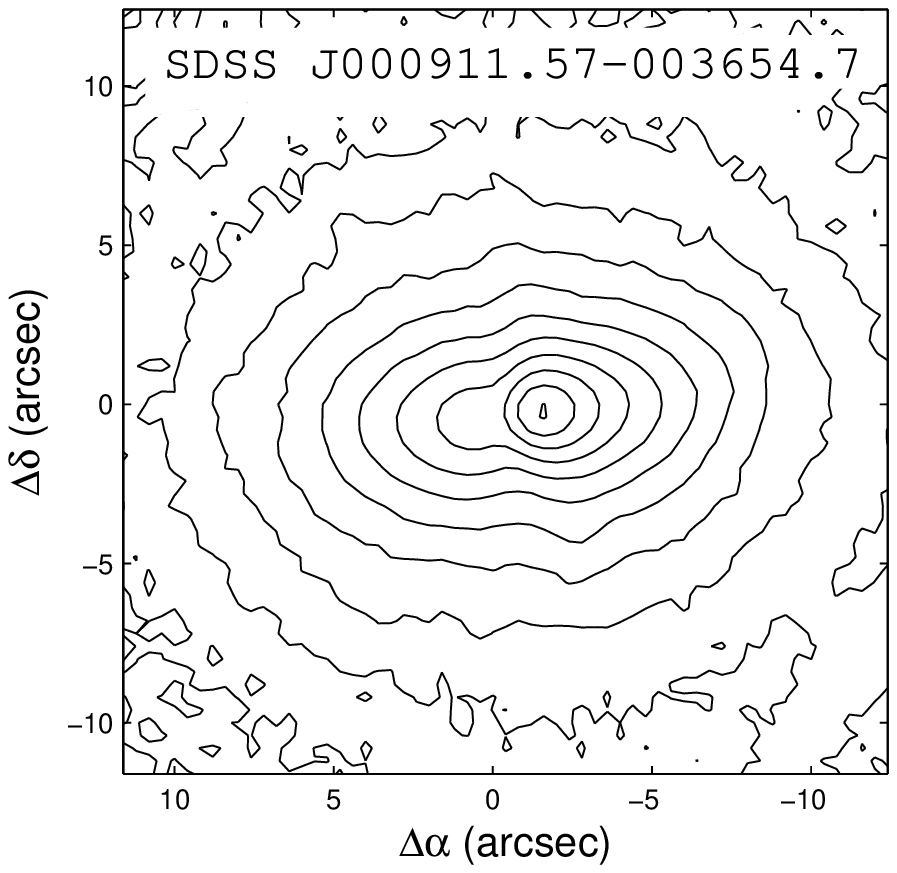}
\end{center}
\caption{SDSS J000911.57-003654.7: a dust-lane elliptical or a PRG?}
 \label{f:DL}
\end{figure}

We emphasize that PRGs host significantly more star formation, but much less nuclear activity, than D-ETGs.
If these two classes of galaxies are formed in similar events, such as by a mass transfer from a low-mass gas-rich system to a high-mass gas-poor system (Wakamatsu 1990;  Whitmore et al.\ 1990; Reshetnikov \& Sotnikova 1997), then differences in the initial conditions and shapes of the gravitational potentials could determine whether the diffuse matter stripped from the donor galaxy would consist also of stars. 
Some unknown mechanisms could also prevent or delay in-situ star formation in the forming polar structure or feed an active nucleus.
In fact, galaxies with faint rings could sometimes be difficult to distinguish from D-ETGs. 
For instance, SDSS J000911.57-003654.7 (SPRC-1) appears at first to be an elliptical galaxy with a dust lane crossing parallel to the minor axis north of the galactic centre. However, closely examining the deep coadded Stripe82 image, shown in Fig.\ \ref{f:DL}, reveals a faint excess of light along the west side of the dust lane which could be starlight arriving from the host galaxy and being scattered by dust, or light emitted directly from stars within a polar structure.

\subsection{Environment}
Mergers are expected to be less common in rich environments due to the high relative velocity of galaxies (Binney \& Tremaine
1987). Therefore, studying the visible neighbourhood of PRGs could provide further clues to the formation of PRGs (Brocca et al.\ 1997).
We searched for nearby companions with similar redshift and brightness and found that the number of galaxies surrounding our PRGs varies from very few in low-density environments to a couple of dozens in the rich environments of clusters. Several PRGs (5, 10 and 14) belong to dense galaxy groups and it is likely that a gravitational link between the galaxies is present. To characterize the environment we visually inspected the SDSS images and spectra of the surrounding galaxies.
PRGs 5, 8 and 14 are surrounded mostly by early-type galaxies. PRGs 3, 9, 11, 15 and 16, which show signatures of recent or on-going star formation, are surrounded by blue star-forming galaxies, implying a gas-rich environment.

To apply an extensive statistical analysis on the environment of PRGs we search for possible group member galaxies among the candidates listed in the SPRC. 
This is done by cross-identifying the `best' and `good' candidates with group members listed in Tempel et al.\ (2011). 
The Tempel et al.\ catalogue, including 78 858 groups, was constructed from SDSS DR8 by applying a group-finding algorithm based on the friends-of-friends method. The catalogue is restricted to the SDSS main area survey, including 576 493 galaxies, and provides information on $\sim$65\% of the SPRC galaxies. To investigate their environments we obtain for each group (or field galaxy) its estimated total luminosity and number of observed members (`richness').
For comparison, we apply the same analysis on the samples of `normal' ETGs, B-ETGs and D-ETGs and present the results in Fig.\ \ref{f:compare}.

Although previous studies argued that the environment of PRGs is similar to that of normal galaxies (Brocca et al.\ 1997), our analysis reveals differences between the various types of ETGs.
In particular, we find that PRGs preferentially inhabit low density environments compared to the other ETG samples. This implies a role of environment which is consistent with mergers as triggers of PRG formation. We note also that D-ETGs are more frequent than PRGs in high density environments. This can be explained by a natural selection effect; in many PRGs the luminous polar structure extends much farther than the main body and could be destroyed more easily by close neighbours.

\begin{figure}
\begin{center}
 \includegraphics[width=8cm]{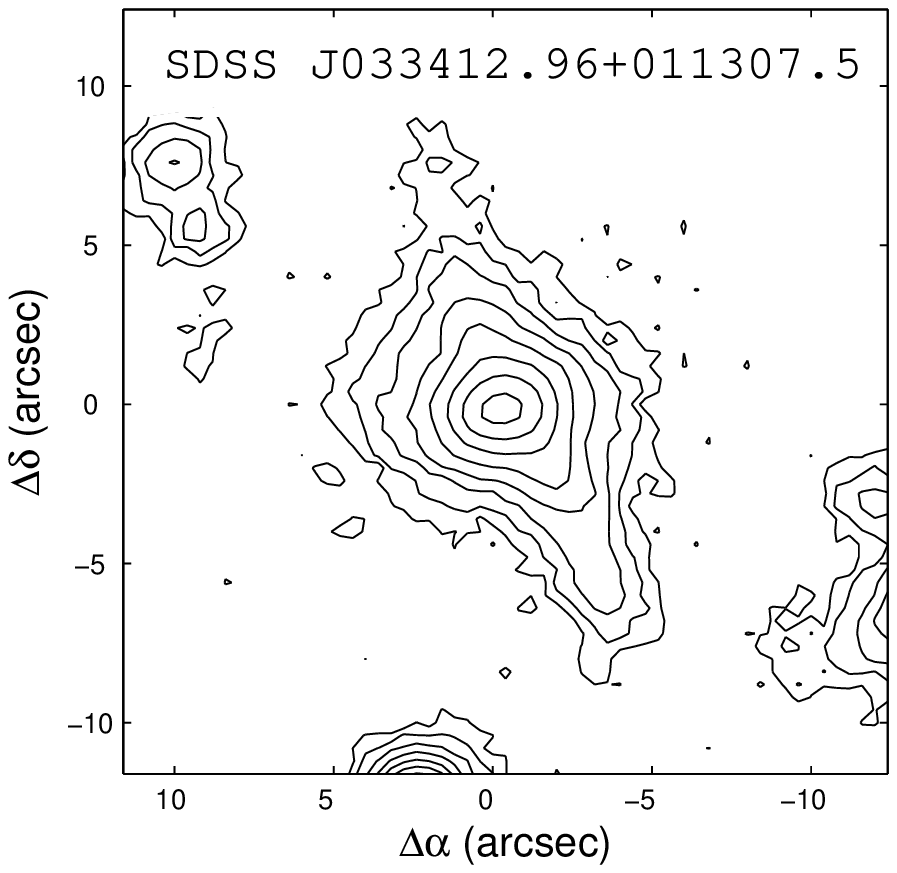}
\end{center}
\caption{SDSS J033412.96+011307.5: a candidate PRG at $z\sim0.3$.}
 \label{f:distant}
\end{figure}

Our study sets the stage for future more detailed work aiming to understand the formation history of
PRGs. 
In particular, we will attempt to confirm that the hosts and rings truly represent two distinct stellar populations by constraining their age and metallicty and determining their kinematics. 
The galaxies in SPRC are typically at higher $cz$ than those published and studied by Whitmore et al.\ (1990) and others, and could bridge the distance gap between nearby PRGs and high-$z$ PRG-candidates found in {\it HST} observations (e.g., Reshetnikov 1997).
Including the coadded Stripe82 images in the Galaxy Zoo project makes it possible to add new promising candidates to the SPRC. SDSS J033412.96+011307.5 at $z\sim0.3$, shown in Fig.\ \ref{f:distant}, is one such example.

\section*{Acknowledgments}
Based on observations with the VATT: the Alice P.\ Lennon Telescope and the Thomas J.\ Bannan Astrophysics Facility.
This study makes use of data from the SDSS (http://www.sdss.org/collaboration/credits.html).

We thank the anonymous referee for constructive comments.
We would like to thank Jonathan Stott S.J.\ for assisting with the VATT observations, Sugata Kaviraj for supplying the sample of dusty early-type galaxies and Benny Trakhtenbrot for his useful comments.
We would also like to thank the many dedicated volunteers of the Galaxy Zoo project and, in particular, the participants of `The Possible Polar Ring Thread' forum.

\end{document}